\documentclass[12pt]{article}
\usepackage{epsfig}
\usepackage{axodraw}
 
\newcommand{\mrm}[1]{\mathrm{#1}}
\newcommand{\as}{\alpha_{\mrm{s}}}
\newcommand{\aem}{\alpha_{\mrm{em}}}
\newcommand{\qbar}{\mrm{\overline{q}}}
\newcommand{\pbar}{\mrm{\overline{p}}}
\newcommand{\mone}[1]{m_\mrm{#1}}
\newcommand{\mtwo}[1]{m_\mrm{#1}^{2}}
\newcommand{\mfour}[1]{m_\mrm{#1}^{4}}
\newcommand{\pt}{p_{\bot}}

\newcommand{\gtrsim}{\raisebox{-0.8mm}%
{\hspace{1mm}$\stackrel{>}{\sim}$\hspace{1mm}}}
\newcommand{\lessim}{\raisebox{-0.8mm}%
{\hspace{1mm}$\stackrel{<}{\sim}$\hspace{1mm}}}
 
\newlength{\abstwidth}
\newlength{\captivewidth}
\newcommand{\captive}[1]{\rule{5mm}{0mm}%
\begin{minipage}{\captivewidth}%
\caption[small]{#1}\end{minipage}}

\begin{document}
 
\sloppy
 
\pagestyle{empty}
 
\begin{flushright}
LU TP 98--9 \\
April 1998
\end{flushright}
 
\vspace{\fill}
 
\begin{center}
{\LARGE\bf A matrix-element-based modification }\\[3mm]
{\LARGE\bf of the parton shower }\\[3mm]
{\Large Gabriela Miu
}\\ [2mm]
{\it Department of Theoretical Physics,}\\[1mm]
{\it Lund University, Lund, Sweden}
\end{center}
 
\vspace{\fill}
\begin{center}
{\bf Abstract}\\[2ex]
\begin{minipage}{\abstwidth}
The transverse momentum distribution of $\mrm{W}^{\pm}$ bosons at hadron colliders is well described by a parton-shower model for small $\pt$ values, but not for large ones. This article is an attempt to give a better description of the distribution by using corrections derived from the matrix-element formalism. The parton shower for $\mrm{q}\qbar'\rightarrow\mrm{W}^{\pm}$ has been modified to resemble the matrix elements of $\mrm{q}\qbar\rightarrow \mrm{gW}$ and $\mrm{qg}\rightarrow \mrm{q'W}$ at large $\pt$ values. Comparisons between different approaches are presented at $\sqrt{s}=1800$ GeV. The results are also compared with experimental data from the D0 collaboration at Fermilab.
\end{minipage}
\end{center}

\vspace{\fill}
 
\clearpage
\pagestyle{plain}
\setcounter{page}{1}
%


%
\section{Introduction}
\label{sec-intro}
%
%
%
The Standard Model is the theory which lies at the heart of modern particle physics. In this theory we distinguish between two kinds of particles: matter particles and gauge bosons. (A boson is a particle of integer spin.) The Standard Model describes the matter particles (quarks and leptons) and their interactions. There are four fundamental forces in nature; interactions between particles take place via the exchange of the gauge bosons corresponding to these forces (see table below).
\begin{equation}
\begin{array}{lll}
\underline{\mrm{Interaction}} & \underline{\mrm{Gauge~bosons}} \\
\mrm{gravitation} & \mrm{graviton}~ (\mrm{spin}=2)\\
\mrm{electromagnetic} & \mrm{\gamma}~ (\mrm{spin}=1) \\
\mrm{weak} & \mrm{W^{+},~~W^{-},~~Z}^{0}~ (\mrm{spin}=1) \\
\mrm{strong} & \mrm{g}_{i},~~i=1 \ldots 8~ (\mrm{spin}=1).
\end{array}
\nonumber
\end{equation}
The gravitational force acts on all forms of energy, but is so weak that it can be disregarded in particle physics. The electromagnetic force acts on all electrical charged particles and is responsible for holding the electrons and the nuclei together in the atoms. It can create and annihilate photons. The weak interaction accounts, amongst other things, for the $\beta$-decay of nuclei. In the Standard Model, the electromagnetic and the weak forces have been unified into one force: the electroweak interaction. The strong force is responsible for holding together the quarks inside the nucleons, as well as the protons and neutrons inside the nuclei. The theory for the colour (strong) interaction is called quantum chromodynamics (QCD)\cite{ref:hadrons}.\\

The quarks and leptons are the fundamental particles that matter is made of. They are all spin-$1/2$ fermions. To each matter particle there corresponds an antiparticle, having the same mass and spin as its partner, but with all other quantum numbers (e.g. electrical charge) having opposite values. \\
The quarks are most significantly affected by the (unified) electroweak force and the strong force. There are six known `flavours' of quarks, which are placed in doublets called `families' or `generations'. Thus, the three generations of quarks are: 
\begin{equation}
\begin{array}{ccc}
\left( \begin{array}{c} \mrm{u} \\ \mrm{d} \end{array} \right) 
& \left( \begin{array}{c} \mrm{c} \\ \mrm{s} \end{array} \right) 
& \left( \begin{array}{c} \mrm{t} \\ \mrm{b} \end{array} \right). 
\end{array}
\nonumber
\end{equation}
The quarks in the upper row all have electric charge $+(2/3)e$, while those in the lower row have charge $-(1/3)e$. From the point of view of the strong interaction, each quark carries one of three possible QCD colour charges: $r$ (red), $g$ (green) or $b$ (blue). \\
All coloured particles are bound inside colourless `hadrons'. From experiments we know that the hadrons are made up not only by quarks, but also by gluons. The concept of partons is thus introduced as a common name for the constituents of the hadrons (cf. the concept of the nucleon in nuclear physics). The hadrons are subdivided into `baryons' (hadrons with half-integer spin) and `mesons' (hadrons with integer spin).\\  
The leptons are unaffected by the strong interaction. They too are placed in doublets and the corresponding three generations are:
\begin{equation}
\begin{array}{ccc}
\left( \begin{array}{c} \nu_{\mrm{e}} \\ \mrm{e} \end{array} \right) 
& \left( \begin{array}{c} \nu_{\mu} \\ \mu \end{array} \right) 
& \left( \begin{array}{c} \nu_{\tau} \\ \tau \end{array} \right). 
\end{array}
\nonumber
\end{equation}
The electron $\mrm{e}$, the muon $\mu$ and the tau $\tau$ all have electric charge $-e$, while the neutrinos $\nu$ are electrically neutral.\\ 

The electromagnetic and the gravitational forces have been known and studied for a long time. When it comes to the strong interaction, since the gluons are confined within hadrons, they can only be studied indirectly. The bosons of the weak interaction are special as they are so massive; they have been much studied since their discovery, in 1983. Until recently, they could only be observed in hadron colliders, but now they have also been produced at LEP2.\\
In this article we focus on the transverse momentum distribution of the $\mrm{W}$ boson in $\mrm{p}\pbar$ collisions. This is related to QCD corrections to the basic electroweak process in which $\mrm{W}$'s are produced. These studies are mainly of interest as a test of our understanding of the QCD, but they implicitly influence our confidence in the Standard Model.\\   
There are two complementary methods of describing the $\pt$ distribution of the $\mrm{W}$: the matrix-element approach and the parton-shower one. The matrix-element formalism has the advantage of being exact (to a given order). Unfortunately, the calculations become more and more complicated as one allows for more and more partons. In the parton-shower formalism, on the other hand, one can describe events with an arbitrary number of partons. This method is approximate, though, and it can only be trusted in the collinear limit (small $\pt$ region).\\
In this Diploma Work we start from the parton-shower formalism and show that it can be extended to hold for large $\pt$ values as well, by adding corrections derived from the matrix-element formalism.\\

This article is organized as follows. Section~2 includes a more detailed description of the matrix-element and the parton-shower formalisms, a presentation of the cross sections for $\mrm{W}$ production and a description of the `backwards evolution' method for reconstructing a parton shower. In Section~3 we present a comparison between the parton-shower and matrix-element differential cross sections, as well as a detailed description of the modeling of the parton shower according to the matrix elements. The results are presented in Section~4. Finally, Section~5 contains a summary of this article.\\
%
%


%
\section{Theory}
\label{sec-theory}
%
%
%
In the simulations of this Diploma Work, protons and antiprotons are allowed to collide at 1800 GeV in the CM reference frame, like it was done in the experiments at Fermilab. One can better understand why such high energies are advantageous to produce a $\mrm{W}$ particle of mass $\approx$ 80 GeV if one remembers that it is actually a quark and an antiquark --- and not the whole proton and antiproton --- that produce the $\mrm{W}$. The quark and the antiquark carry only a fraction of the energies of the proton and the antiproton. \\

In QED it is well known that an electrically charged particle which is accelerating emits photons, for instance as is the case when an electron moves within a magnetic field. In general, high-energy interactions between charged particles are accompanied by photon emission. This phenomenon is known as bremsstrahlung. \\
In a similar way, in QCD, scatterings between particles can give rise to new partons being produced. Unfortunately ambiguities arise when attempts are made to distinguish between bremsstrahlung and hard-scattering partons. In a general process $ab\rightarrow cd$, partons can be radiated both by the incoming $a$ and $b$ (initial-state radiation), as well as by the outgoing $c$ and $d$ (final-state radiation) partons. Furthermore, interference between the initial- and the final-state radiation can arise, and these effects are not negligible. To simplify matters one can choose to consider processes where, for instance, only initial-state radiation is possible, such as $\mrm{q} \qbar' \rightarrow \mrm{W}^{+} \rightarrow \mrm{e}^{+}\nu_{\mrm{e}}$ or $\mrm{q}\qbar \rightarrow \mrm{Z}^{0} \rightarrow \mrm{e}^{+} \mrm{e}^{-}$. It is this initial-state radiation that is responsible for the transverse momentum spectra of the $\mrm{W}$ and $\mrm{Z}$. (Actually, the shower initiators also possess some transverse momentum because they were confined within the protons. The effects that these so-called primordial $\pt$ have on the spectra are small, and will not be considered here.)\\

In this article we study the transverse momentum distribution of the $\mrm{W}$ bosons.
The existence of the $\mrm{W}$ particle was first proved in 1983 at CERN, where very energetic protons and antiprotons were allowed to collide. The $\mrm{W}$ particle is unstable since it can decay into less massive states. Thus it has a characteristic lifetime $\tau$ and, because of the Heisenberg uncertainty principle, an uncertainty in the mass.  As a result, it is possible for $\mrm{W}$'s of masses somewhat different from the nominal one to be produced.                      
%
%


%
\subsection{Mandelstam variables}
\label{subsec-Mandelstam}
\begin{figure}[t]
\begin{center}  
\begin{picture}(280,140)(0,0)
\Line(20,100)(65,100)
\Text(10,100)[]{$A$}
\Oval(70,100)(15,5)(0)
\Line(20,20)(65,20)
\Text(10,20)[]{$B$}
\Oval(70,20)(15,5)(0)
\Line(70,110)(250,110)
\Line(70,100)(250,100)
\Line(70,90)(110,75)
\Text(100,75)[t]{$a$}
\Line(70,10)(250,10)
\Line(70,20)(250,20)
\Line(70,30)(110,45)
\Text(100,45)[b]{$b$}
\Line(110,75)(110,45)
\Vertex(110,75){2}
\Vertex(110,45){2}
\Oval(110,60)(25,25)(0)
\Photon(110,75)(250,130){3}{10}
\Text(120,75)[t]{$c$}
\Gluon(110,45)(250,30){2.5}{15}
\Text(120,50)[b]{$d$}
\Line(250,115)(255,115)
\Line(255,115)(255,5)
\Line(255,5)(250,5)
\Text(265,130)[]{$C$}
\Text(265,60)[]{$D$}
\end{picture}
\end{center}
\caption{\label{fig:process-subprocess} The process $A+B\rightarrow C+D$. The
circled region marks the corresponding subprocess $a+b\rightarrow c+d$.}
\end{figure}
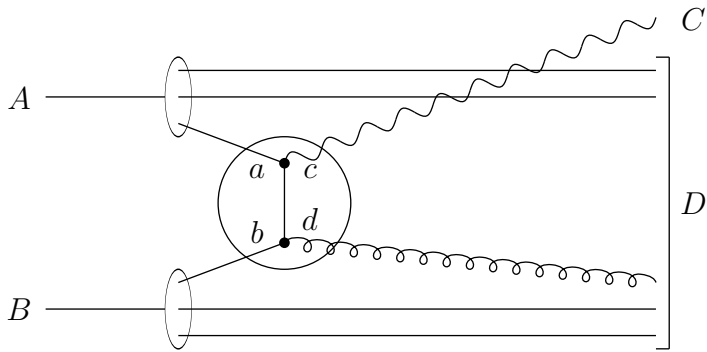

%
%
In a general process of the form $A+B\rightarrow C+D$ the following invariants
--- so-called Mandelstam variables --- can be defined
\begin{eqnarray} 
s &=& (p_{A}+p_{B})^2,
\\
t &=& (p_{A}-p_{C})^2,
\\
u &=& (p_{A}-p_{D})^2,
\end{eqnarray}
where $p_{i}$ is the four-momentum of particle $i$. The variable $s$ is
simply the square of the total energy in the CM system. If $m_{i}$ represents the mass of particle $i$ then we obtain the relation
\begin{equation}
s+t+u = \sum_{i} m_{i}^{2}.
\label{s+t+u} 
\end{equation}
If one actually knows the subprocess $a+b\rightarrow c+d$, one can define the corresponding $\hat{s}$, $\hat{t}$ and $\hat{u}$ by using $p_{a}$, $p_{b}$, $p_{c}$ and $p_{d}$ as arguments (see figure \ref{fig:process-subprocess}).
For instance if $A$ is a proton and $B$ an antiproton then $a$ can be a quark and $b$ an antiquark.\\
One can define some dimensionless variables $x_{i}$ according to
\begin{eqnarray} 
p_{a} &=& x_{1}p_{A},
\\
p_{b} &=& x_{2}p_{B},
\end{eqnarray}
relating the incoming four-momenta of the process and the subprocess. This can be done if one assumes that $a$ ($b$) moves along the direction of $A$ ($B$), which is a good approximation at high energies.
In order to derive a useful relation between $s$ and $\hat{s}$, consider the process $\mrm{p}\pbar\rightarrow \mrm{WX}$ with the corresponding subprocess $\mrm{q}\qbar'\rightarrow \mrm{W}$. We have
\begin{eqnarray}
s &=& (p_{\mrm{p}}+p_{\pbar})^2,
\\
\hat{s} &=& (p_{\mrm{q}}+p_{\qbar})^2,
\end{eqnarray}
or
\begin{eqnarray}
s &=& 2 \mtwo{p} + 2 p_{\mrm{p}} p_{\pbar},
\label{s_temp}
\\
\hat{s} &=& (x_{1} p_{\mrm{p}} + x_{2} p_{\pbar})^2 = (x_{1}^{2}+x_{2}^{2})
\mtwo{p} + 2x_{1}x_{2} p_{\mrm{p}} p_{\pbar}.
\label{shat_temp}
\end{eqnarray}
By noticing that $s > \hat{s}$ and $\hat{s}\approx\mtwo{W} \gg \mtwo{p}$,
one obtains from equations (\ref{s_temp}) and (\ref{shat_temp})
\begin{equation}
\hat{s} \approx x_{1}x_{2}s. 
\end{equation}
Alternatively, it is possible to take the equation
\begin{equation}
\hat{s} =  x_{1}x_{2}s 
\end{equation}
as a starting point and use it in the definition of $x_{1}$ and $x_{2}$.
%
%


%
\subsection{Matrix elements versus parton showers}
\label{subsect-ME-PS}
%
%
In order to describe interacting fields, in which particles can be scattered, created and annihilated, one has to solve the very difficult equations which arise. Considering in particular QED and QCD (which both have gauge bosons that are massless), one approach is to work within the framework of perturbation theory. Technically this means that the Hamiltonian of the system is divided into a free-field term plus an interaction term. For a sufficiently weak interaction, this last term can be treated as a perturbation.\\

A matrix element expresses the probability amplitude for a scattering process
\begin{equation}
\mathcal{M}=\langle f|\mrm{S}|i\rangle. 
\end{equation}
Here $|i\rangle$ represents the initial state long before the scattering occurs, specifying a definite number of particles and their properties when they are far apart from each other. Similarly $|f\rangle$ describes the final state, after the particles have interacted and when they are far apart again. $\mrm{S}$ is the so-called scattering matrix (S-matrix) \cite{ref:matrix}.\\
Once the initial state and the final state have been fixed, the cross section is obtained by summing and then squaring all possible Feynman diagrams that contribute to the process, in the usual quantum mechanical manner. The more vertices we include in the diagrams, the more difficult the calculations become.\\
To calculate the contribution from each diagram one makes use of the Feynman rules. A factor of $\sqrt{\aem}$ ($\sqrt{\as}$) corresponds to each QED (QCD) three-particle vertex. The four-gluon vertex of QCD corresponds to a $\as$ factor and is thus more suppressed.\\
The cross section can be expanded in a $\aem$ ($\as$) series. In the case of QED, the expansion terms decrease at a relatively fast rate. Consequently, it is a good approximation to keep only the first few terms of the expansion, as the rest terms are negligible. For QCD, the terms do not decrease as fast, and more terms in the expansion are of significant size.\\ 
In QED the cross section is expanded in terms of the small dimensionless fine-structure constant $\aem$, which measures the strength of the coupling between the electron and the photon 
\begin{equation}
\aem = \frac{e^{2}}{4\pi \varepsilon_{0} \hbar c}\approx
\frac{1}{137}. 
\end{equation}
For QCD one can take a similar approach by expanding in the corresponding so called strong-coupling constant. This in not really constant, but is 'running' with the energy. (At this point we should add that in QED, $\aem$ is not a constant either; the expression given above is the value at $Q^2=0$.) To first order, it is given by 
\begin{equation}
\as(Q^2) = \frac{12\pi}{(33-2 n_{\mrm{f}}) 
\ln(Q^{2}/\Lambda^{2}_{\mrm{QCD}})}, 
\end{equation}
where $Q$ is the energy scale for the process, $n_{\mrm{f}}$ is the number of quark flavours available at the actual energy (usually 4--5) and $\Lambda_{\mrm{QCD}}\sim$ 200 MeV is the QCD scale parameter \cite{ref:alpha-strong}.\\
The matrix-element description has the advantage of being exact (to a given order). It can be used successfully for processes where the outgoing particles are well separated. Unfortunately it is not so easy to apply in the collinear limit, since $\as$ becomes larger and the series converges slower. Another drawback of the matrix-element description is that the calculations become complicated very fast as one allows for more photons or gluons in the processes. Also, there are uncertainties in the choice of the $Q^{2}$ scale for $\as$.\\ 

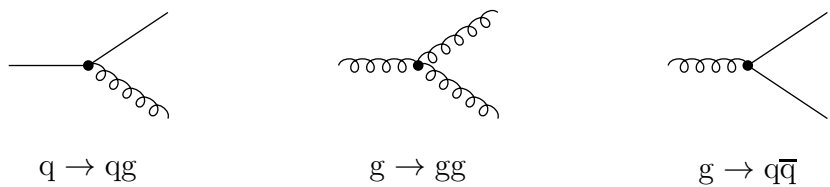
\begin{figure}[htb]
\begin{center}
\begin{picture}(60,80)(0,0)
\Line(0,50)(30,50)
\Vertex(30,50){2}
\Line(30,50)(60,70)
\Gluon(30,50)(60,30){2.5}{5}
\Text(30,10)[]{$\mrm{q}\rightarrow \mrm{qg}$}
\end{picture}
\hspace{2cm}
\begin{picture}(60,80)(0,0)
\Gluon(0,50)(30,50){2.5}{4}
\Vertex(30,50){2}
\Gluon(30,50)(60,70){2.5}{5}
\Gluon(30,50)(60,30){2.5}{5}
\Text(30,10)[]{$\mrm{g}\rightarrow \mrm{gg}$}
\end{picture}
\hspace{2cm}
\begin{picture}(60,80)(0,0)
\Gluon(0,50)(30,50){2.5}{4}
\Vertex(30,50){2}
\Line(30,50)(60,70)
\Line(30,50)(60,30)
\Text(30,10)[]{$\mrm{g}\rightarrow \mrm{q\overline{q}}$}
\end{picture}
\end{center}
\caption{\label{fig:QCD-branchings}The basic QCD branchings}
\end{figure}

A complementary approach to the matrix-element description is the parton-shower one. The building blocks of the parton shower are branchings of the form $a\rightarrow bc$. These can be repeated, forming a tree-like structure. 
Each branching vertex is associated with some relative transverse and longitudinal momentum between the partons $b$ and $c$. For an $a\rightarrow bc$  branching, we let the variable $z$ represent the fraction of the longitudinal momentum of the incoming particle $a$ that particle $b$ is taking. \\
The three basic QCD branchings $\mrm{q}\rightarrow\mrm{qg}$, $\mrm{g}\rightarrow\mrm{gg}$ and $\mrm{g}\rightarrow\mrm{q}\qbar$ are shown in figure \ref{fig:QCD-branchings} (the branchings in the QED case are analogous, except that there is no three--photon vertex, since photons do not interact with each other). 
The branching probability $P_{a\rightarrow bc}(z)$ for each case is given by the so-called Altarelli-Parisi (AP) splitting kernels   
\begin{eqnarray}
P_{\mrm{q}\rightarrow\mrm{qg}}(z) &=& \frac{4}{3} \, \frac{1+z^2}{1-z},
\label{P1}
\\
P_{\mrm{g}\rightarrow\mrm{gg}}(z) &=& 3 \, \frac{(1-z(1-z))^2}{z(1-z)},
\label{P2}
\\
P_{\mrm{g}\rightarrow\mrm{q}\qbar}(z) &=& n_{\mrm{f}} \, \frac{1}{2} \, 
(z^2+(1-z)^2).
\label{P3}
\end{eqnarray}
These can be derived from the expressions of 3-parton-matrix-elements cross sections, by taking the limit of two partons being collinear. We will return to this in Section~\ref{subsec-comparing}. Notice that the second equation corresponds to two identical particles in the final state. The factor 3 therefore has to be replaced by 6 in the case that one is asking for the probability of finding a gluon in the final state, given that its mother was a gluon. The first expression is singular in the limit $z\rightarrow 1$, which corresponds to a `soft' (of low energy) gluon being emitted. Similarly, the second expression is singular both as $z\rightarrow 1$ and as $z\rightarrow 0$ corresponding to any one of the two emitted gluons being `soft'.\\
In the described model of the parton shower, the different branchings are regarded as independent. As a consequence the total amount of evolution is overestimated. In the case of initial-state radiation, coherence effects can be taken into account by demanding that the virtualities associated with the partons on the `main chain' -- the chain that starts with the initiating parton and ends with the scattered one -- are ordered, with the largest being closest to the hard scattering (cf. Section~\ref{subsec-back-ev}). \\
The parton-shower description has the advantage that one can allow for an arbitrary number of particles both in the initial and the final state. Also, it gives a good description in the collinear limit. Unfortunately the description is approximate.\\

There are two alternative approaches one can adopt in describing the change in the number of partons, as a function of the resolution. The first picture is an exclusive one: we follow the original parton as it repeatedly branches. We will return to this in Section~\ref{subsec-back-ev}. The second approach is the parton-density picture: we give an inclusive description of the number of partons of a certain kind. Consider the proton, for instance. Nonperturbatively, it is a bound state and its parton composition is not known beforehand. Given some (fitted) distribution at a low $Q^2$ scale, the change in the number of partons $b$ is given by the so-called DGLAP evolution equations 
\begin{equation}
\frac{\mrm{d}f_{b}(x,Q^2)}{\mrm{d}Q^2} = \sum_{a} \int_{x}^{1} \frac{\mrm{d}x'}{x'} \frac{1}{Q^2} \, \frac{\as(Q^2)}{2\pi} \,  f_{a}(x',Q^2) \, 
P_{a\rightarrow bc}(z).
\label{APQ}
\end{equation}
This inclusive picture of showers is also used in the matrix-element formalism, but, unlike in the normal showers, no $\pt$ is assigned to the partons. If we put $t=\ln(Q^{2}/\Lambda^{2}_{\mrm{QCD}})$, the above equation can be rewritten as
\begin{equation}
\frac{\mrm{d}f_{b}(x,t)}{\mrm{d}t} = \sum_{a} \int_{x}^{1} \frac{\mrm{d}x'}
{x'} \, \frac{\as(t)}{2\pi} \, f_{a}(x',t) \, P_{a\rightarrow bc}(z).
\label{APt}
\end{equation}
The functions $f_{i}(x,t)$ are the so-called parton distributions, which measure the probability of finding a parton $i$ carrying a fraction $x$ of the total momentum of the incoming hadron. The variable $Q^2$ represents the scale for the process: the higher the $Q^2$ value, the finer a structure in the hadron one can distinguish. Here $x'$ and $zx'=x$ are the momentum fractions that the partons $a$ and $b$, respectively, take from the total hadron momentum, with $z$ defined as before (see figure \ref{fig:x=zx'}).
\begin{figure}[htb]
\begin{center}
\begin{picture}(160,80)(0,0)
\Line(20,40)(70,40)
\Text(45,45)[]{$a$}
\Text(10,40)[]{$x'$}
\Vertex(70,40){2}
\Gluon(70,40)(120,60){3}{6}
\Text(100,62)[]{$c$}
\Text(130,60)[lb]{$x'-x$}
\Line(70,40)(120,20)
\Text(100,35)[]{$b$}
\Text(130,20)[lb]{$x=zx'$}
\end{picture}
\end{center}
\caption{\label{fig:x=zx'} The sharing of the momentum fractions $x'$, $x$ and $z$.}
\end{figure}
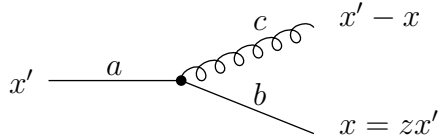\\
To better understand the meaning of the DGLAP evolution equation, consider the case when $b$ represents a quark $\mrm{q}$. Then equation (\ref{APt}) becomes 
\begin{equation}
\frac{\mrm{d}f_{\mrm{q}}(x,t)}{\mrm{d}t} = \int_{x}^{1} \frac{\mrm{d}x'}{x'} \,
\frac{\as(t)}{2\pi} \, f_{\mrm{q}}(x',t) \, P_{\mrm{q}\rightarrow\mrm{qg}}(z) 
+\int_{x}^{1} \frac{\mrm{d}x'}{x'} \, \frac{\as(t)}{2\pi} \, f_{\mrm{g}}(x',t) \, P_{\mrm{g}\rightarrow\mrm{q}\qbar}(z).
\label{APq}
\end{equation}
The first term in (\ref{APq}) expresses the fact that the quark with momentum fraction $x$ could have originated from a parent quark with a larger momentum fraction $x'$ which has emitted a gluon. The probability for the emission is proportional to $\as P_{\mrm{q}\rightarrow\mrm{qg}}(z=\frac{x}{x'})$ and one has to integrate over all possible momentum fractions $x'>x$ of the parent. The second term can be interpreted in a similar way, but now the parent is a gluon.\\

To summarize, we conclude that the matrix-element description is useful for the description of the $\pt$ spectrum at large values (when $\as$ is small), while the parton-shower description is a good candidate in the small $\pt$ region (collinear limit).

%
%


\subsection{Cross sections for $\mrm{W}$ production}
\label{subsec-cross-sect}
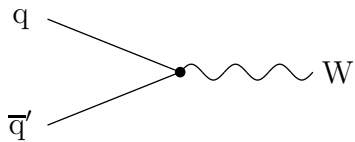
\begin{figure}[b]
\begin{center}
\begin{picture}(160,60)(0,0)
\Line(20,50)(70,30)
\Text(10,50)[]{$\mrm{q}$}
\Line(20,10)(70,30)
\Text(10,10)[]{$\qbar'$}
\Vertex(70,30){2}
\Photon(70,30)(120,30){3}{3}
\Text(130,30)[]{$\mrm{W}$}
\end{picture}
\end{center}
\caption{\label{fig:qqbar->W} $\mrm{q}\qbar' \rightarrow \mrm{W}$.}
\end{figure}

%
%
%
In a collider, the number of events of a particular kind is given by
\begin{equation}
N=\sigma L 
\end{equation}
where the integrated luminousity $L$ is a characteristic of the accelerator and the total cross section $\sigma$ is a property of the process in question.\\
If protons and antiprotons are allowed to collide at the CM energy $\sqrt{s}$, then the lowest-order process that produces $\mrm{W}$'s is $\mrm{q}\qbar' \rightarrow \mrm{W}$ (see figure \ref{fig:qqbar->W}). 
The total cross section is given by the expression
\begin{equation} 
\sigma = \sigma_{0}\int\!\int \mrm{d}x_{1}\,\mrm{d}x_{2} \ F(x_{1},x_{2}) \,   
\delta\left(x_{1}x_{2}-\frac{\mtwo{W}}{s}\right),
\label{sigma1}
\end{equation}
where
\begin{eqnarray}
\sigma_{0} &=& \frac{\pi^{2}}{3} \, \frac{\aem}{\sin^{2}\theta_{\mrm{W}}} \,
\frac{1}{\mtwo{W}},
\label{sigma0}
\\
F(x_{1},x_{2}) &=&\sum_{\mrm{q}\qbar'} \{x_{1} f^{\mrm{p}}_{\mrm{q}}(x_{1},Q^2)
x_{2}f^{\pbar}_{\qbar}(x_{2},Q^2)+x_{1} f^{\mrm{p}}_{\qbar}(x_{1},Q^2)
x_{2}f^{\pbar}_{\mrm{q}}(x_{2},Q^2)\}|\mrm{V}^{\mrm{CKM}}_{\mrm{q}\qbar'}|^{2}.
\label{F}
\end{eqnarray}
$\mrm{V}^{\mrm{CKM}}$ is the Cabibbo-Kobayashi-Maskawa (CKM) matrix. It is a 
$3\times 3$ unitary matrix containing three independent mixing angles and one phase. The angles are generalizations of the Cabibbo angle, describing the possible mixings between the (u,d), (c,s) and (t,b) doublets. As an example, one would naively expect a process like $\mrm{u}\mrm{\overline{d}}\rightarrow\mrm{W}$ to be allowed but $\mrm{u}\mrm{\overline{s}}\rightarrow\mrm{W}$ to be forbidden. Because of the mixing, however, even the latter process is possible, although at a reduced rate.\\
The mass-distribution spectrum for the $\mrm{W}$ has a Breit-Wigner shape, with a peak at the nominal mass $\mrm{m_{W}}\approx$ 80 GeV. 
The Breit-Wigner distribution is given by
\begin{equation}
\mrm{BW}(\hat{s}) = \frac{1}{\pi} \, \frac{\mone{W} \mrm{\Gamma}_{\mrm{W}}}
{(\hat{s}-\mtwo{W})^{2}+\mtwo{W} \mrm{\Gamma}_{\mrm{W}}^{2}}, 
\nonumber
\end{equation}
where $\Gamma=1/\tau$ is the full width at half maximum. It expresses the fact that there is a non-zero probability of producing $\mrm{W}$'s with the `wrong' mass. \\
Unless one is explicitly interested in the $\mrm{W}$ mass distribution, it is
a good approximation to replace the actual Breit-Wigner resonance by a $\delta$-function, as we did in equation~(\ref{sigma1}).\\

\begin{figure}[h]
\begin{center}
\begin{picture}(140,100)(0,0)
\Line(30,80)(70,65)
\Text(5,80)[]{1}
\Text(20,80)[t]{$\mrm{q}$}
\Line(30,20)(70,35)
\Text(5,20)[]{2}
\Text(20,20)[]{$\qbar'$}
\Line(70,65)(70,35)
\Vertex(70,65){2}
\Vertex(70,35){2}
\Gluon(70,65)(110,80){2.5}{5}
\Text(120,80)[t]{$\mrm{g}$}
\Text(135,80)[]{3}
\Photon(70,35)(110,20){3}{3}
\Text(120,20)[]{$\mrm{W}$}
\Text(135,20)[]{4}
\Text(70,0)[]{$t$-channel graph}
\end{picture}
\hspace{2cm}
\begin{picture}(140,100)(0,0)
\Line(30,80)(70,65)
\Text(5,80)[]{1}
\Text(20,80)[t]{$\mrm{q}$}
\Line(30,20)(70,35)
\Text(5,20)[]{2}
\Text(20,20)[]{$\qbar'$}
\Line(70,65)(70,35)
\Vertex(70,65){2}
\Vertex(70,35){2}
\Gluon(70,35)(110,80){2.5}{7}
\Text(120,80)[t]{$\mrm{g}$}
\Text(135,80)[]{3}
\Photon(70,65)(110,20){3}{5}
\Text(120,20)[]{$\mrm{W}$}
\Text(135,20)[]{4}
\Text(70,0)[]{$u$-channel graph}
\end{picture}
\end{center}
\caption{\label{fig:qqbar->gW} $\mrm{q}\qbar'\rightarrow \mrm{gW}$}
\end{figure}
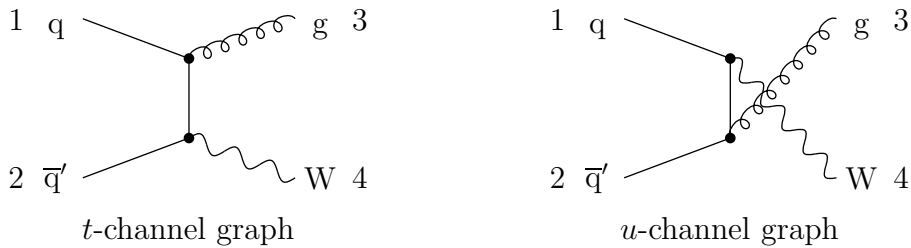

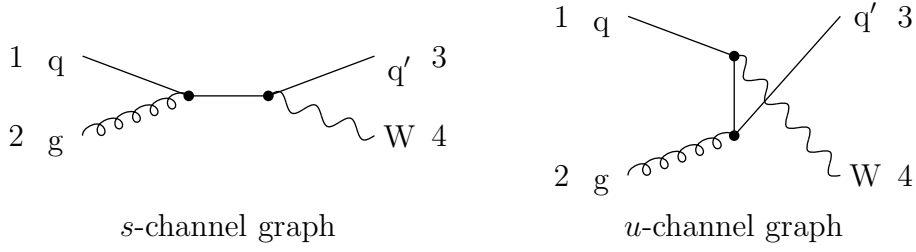
\begin{figure}[h]
\begin{center}
\begin{picture}(170,100)(0,0)
\Line(30,65)(70,50)
\Text(5,65)[]{1}
\Text(20,65)[t]{$\mrm{q}$}
\Gluon(30,35)(70,50){2.5}{5}
\Text(5,35)[]{2}
\Text(20,35)[t]{$\mrm{g}$}
\Vertex(70,50){2}
\Line(70,50)(100,50)
\Line(100,50)(140,65)
\Text(150,65)[t]{$\mrm{q}'$}
\Text(165,65)[]{3}
\Photon(100,50)(140,35){3}{3}
\Text(150,35)[]{$\mrm{W}$}
\Text(165,35)[]{4}
\Vertex(100,50){2}
\Text(85,0)[]{$s$-channel graph}
\end{picture}
\hspace{1cm}
\begin{picture}(140,100)(0,0)
\Line(30,80)(70,65)
\Text(5,80)[]{1}
\Text(20,80)[t]{$\mrm{q}$}
\Gluon(30,20)(70,35){2.5}{5}
\Text(5,20)[]{2}
\Text(20,20)[t]{$\mrm{g}$}
\Line(70,65)(70,35)
\Vertex(70,65){2}
\Vertex(70,35){2}
\Line(70,35)(110,80)
\Text(120,80)[]{$\mrm{q}'$}
\Text(135,80)[]{3}
\Photon(70,65)(110,20){3}{5}
\Text(120,20)[]{$\mrm{W}$}
\Text(135,20)[]{4}
\Text(70,0)[]{$u$-channel graph}
\end{picture}
\end{center}
\caption{\label{fig:qg->qW} $\mrm{qg}\rightarrow \mrm{q^{'}W}$}
\end{figure}

In the collinear limit, processes like $\mrm{q}\qbar'\rightarrow \mrm{gW}$ and $\mrm{qg}\rightarrow \mrm{q'W}$ are already included in the expression for the cross section given in equation~(\ref{sigma1}). (The $Q^2$ dependence of the parton distributions $f_{i}(x,t)$ from equation~(\ref{F}) implicitly means that the $x$ scaling violations are included.) Here we look at the corresponding matrix elements, relevant at large $\pt$ in particular. 
For the $\mrm{q}\qbar'\rightarrow \mrm{gW}$ process, the contributing Feynman diagrams are shown in figure \ref{fig:qqbar->gW}.
These diagrams interfere and the cross section is given by
\begin{eqnarray}
\sigma &=& \int\!\int\!\int \mrm{d}x_{1} \, \mrm{d}x_{2} \, \mrm{d}\hat{t} \,
\frac{F(x_{1},x_{2})}{x_{1}x_{2}} \, \frac{\mrm{d}\hat{\sigma}}
{\mrm{d}\hat{t}}, 
\label{sigma2}
\\
\mrm{where~~}
\frac{\mrm{d}\hat{\sigma}}{\mrm{d}\hat{t}} &=& \sigma_{0} \, \frac{4}{3} \,
\frac{\as}{2\pi} \, \frac{\mtwo{W}}{\hat{s}^2} \, \frac{\hat{t}^2+\hat{u}^2+
2\mtwo{W}\hat{s}}{\hat{t}\hat{u}},
\label{diff.c.s.qqbar}
\end{eqnarray}
and where $\sigma_{0}$ and $F(x_{1},x_{2})$ are as in equations (\ref{sigma0})
and (\ref{F}). Notice that $\mrm{d}\hat{\sigma} / \mrm{d}\hat{t}$ is 
symmetric in $\hat{t}$ and $\hat{u}$.\\ 
In the case of $\mrm{qg}\rightarrow \mrm{q^{'}W}$ the contributing diagrams are shown in figure \ref{fig:qg->qW}.   
The cross section is formally given by the same expression as for the 
$\mrm{q}\qbar'\rightarrow \mrm{gW}$ case (equation (\ref{sigma2})). Because we here also have a gluon in the initial state, the terms in the expression for $F(x_{1},x_{2})$ will include structure functions of the type $f^{\mrm{p}}_{\mrm{g}}(x,Q^2)$. Also, $\mrm{d}\hat{\sigma} / \mrm{d}\hat{t}$ is here given by
\begin{eqnarray}
\frac{\mrm{d}\hat{\sigma}}{\mrm{d}\hat{t}} &=& \sigma_{0} \, \frac{1}{2} \,
\frac{\as}{2\pi} \, \frac{\mtwo{W}}{\hat{s}^2} \, \frac{\hat{s}^2+\hat{u}^2+
2\mtwo{W}\hat{t}}{(-\hat{s})\hat{u}}.
\label{diff.c.s.qg}
\end{eqnarray}
Notice that $\mrm{d}\hat{\sigma} / \mrm{d}\hat{t}$ is now symmetric in $-\hat{s}$ and $\hat{u}$. (With a different notation for the particles in figure \ref{fig:qg->qW}, i.e. if particles 3 and 4 are interchanged, the $\hat{t}$ and $\hat{u}$ would have been interchanged in the expression $\mrm{d}\hat{\sigma} / \mrm{d}\hat{t}$ above).\\

The $\mrm{W}^{+}$ can decay through one of the following main channels:
\begin{equation}
\mrm{W}^{+}\rightarrow\mrm{e}^{+}\nu_{\mrm{e}},\mrm{u}\mrm{\overline{d}},
\mrm{\mu}^{+}\nu_{\mrm{\mu}},\mrm{c}\mrm{\overline{s}},
\mrm{\tau}^{+}\nu_{\mrm{\tau}},\mrm{t}\mrm{\overline{b}},
\nonumber
\end{equation}
and the $\mrm{W}^{-}$ decays in a similar way. (Actually, more channels are possible because of the mixing effect mentioned above.) In reality the $\mrm{t}\mrm{\overline{b}}$ channel is not allowed, since $\mrm{W}$ is not heavy enough. Because quarks can have three different colours, one has to remember that a decay like $\mrm{W}^{+}\rightarrow\mrm{u}\mrm{\overline{d}}$ actually corresponds to three different channels 
$\mrm{W}^{+}\rightarrow\mrm{u}_{r} \mrm{\overline{d}}_{r}$, 
$\mrm{W}^{+}\rightarrow\mrm{u}_{g} \mrm{\overline{d}}_{g}$ or
$\mrm{W}^{+}\rightarrow\mrm{u}_{b} \mrm{\overline{d}}_{b}$.\\
The probability of $\mrm{W}$ to decay to a final state $i$ is given by $\Gamma^{i}_{\mrm{W}}/\Gamma^{\mrm{tot}}_{\mrm{W}}$, where
\begin{equation}
\Gamma^{i}_{\mrm{W}} = \frac{1}{12} \, \frac{\aem}{\sin^{2}\theta_{\mrm{W}}} \,
(\mrm{N}_{\mrm{c}})_{i} \, |\mrm{V}^{\mrm{CKM}}_{i}|^{2} \,\mone{W}.
\end{equation}
$(\mrm{N}_{\mrm{c}})_{i}$ is a colour factor which equals 3 for quarks and
1 for leptons. $\Gamma^{\mrm{tot}}_{\mrm{W}}$ is the total width for the $\mrm{W}$ decay
\begin{equation}
\Gamma^{\mrm{tot}}_{\mrm{W}} = \sum_{i} \Gamma^{i}_{\mrm{W}} 
\approx2\,\mrm{GeV}.
\end{equation}
Thus, one obtains for the lifetime of the $\mrm{W}$ 
\begin{equation}
\tau=\hbar/\Gamma\sim{10^{-25}} \, \mrm{s}.\\ 
\end{equation}
%
%


%
\subsection{Backwards Evolution}
\label{subsec-back-ev}
%

\begin{figure}[ht]
\begin{center}
\begin{picture}(360,190)(0,0)
\Text(30,170)[lb]{initial-state parton shower}
\Text(330,170)[rb]{final-state parton shower}
\Line(50,150)(150,100)
\Gluon(80,135)(110,150){2.5}{5}
\Vertex(80,135){2}
\Gluon(110,120)(140,135){2.5}{5}
\Vertex(110,120){2}
\Line(50,50)(150,100)
\Gluon(90,70)(120,50){2.5}{5}
\Vertex(90,70){2}
\Gluon(120,85)(150,70){2.5}{5}
\Vertex(120,85){2}
\Vertex(150,100){2}
\Photon(150,100)(210,100){3}{3}
\Text(180,115)[]{$\mrm{W}$}
\Vertex(210,100){2}
\Line(210,100)(310,150)
\Gluon(240,115)(270,100){2.5}{5}
\Vertex(240,115){2}
\Gluon(270,130)(300,120){2.5}{5}
\Vertex(270,130){2}
\Line(210,100)(310,50)
\Gluon(270,70)(290,80){2.5}{3}
\Vertex(270,70){2}
\Gluon(290,80)(320,87){2.5}{4}
\Gluon(290,80)(320,75){2.5}{4}
\Vertex(290,80){2}
\Text(50,25)[]{$Q_{0}^2$}
\Text(180,25)[]{$Q_{\mrm{max}}^2 \!= \! \mtwo{W}$}
\Text(310,25)[]{$m^{2}_{0}$}
\LongArrow(70,25)(140,25)
\LongArrow(290,25)(220,25)
\end{picture}
\end{center}
\caption{\label{fig:initial-final-ps} Initial-state and final-state parton showers. In both cases the virtualities are increasing towards the hard scattering.}
\end{figure}
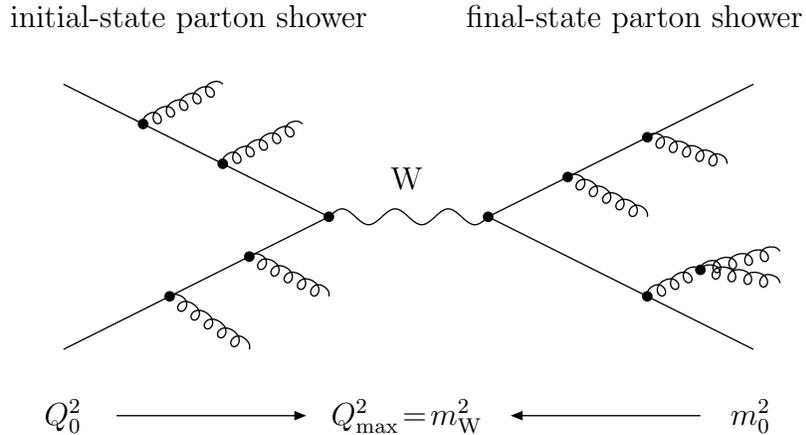

In the QCD shower description, the process $\mrm{q}\qbar'\rightarrow \mrm{W}$
can look schematically as in figure~\ref{fig:initial-final-ps}. Parton showers are allowed in both the initial and final state. The hard process (highest virtuality) is the part of the diagram where the $\mrm{W}$ is produced; it is the propagator that sets the scale for the process. The maximum virtuality is set equal to the square of the mass of the propagator, $Q_{\mrm{max}}^2 = \mtwo{W}$. Both in the initial- and final-state cascade the virtuality increases towards the hard process.\\
In this article we are interested in the transverse momentum for the $\mrm{W}$'s produced, so only the initial-state parton shower is relevant.\\
%
%
In describing the initial-state radiation, one speaks of a `main chain' of consecutive branchings, which connects the initiating parton to the scattered one. In the successive branchings $a\rightarrow bc$ of the cascade, the partons $a$ and $b$ are on the main chain. Conventionally, these partons are assigned a spacelike virtuality, while $c$ is assumed to be on mass shell (or of timelike virtuality, in which case it will initiate a final-state cascade). Since the (spacelike) virtuality of the cascade is carried by one single parton at a time, it is possible to equate this with the $Q^2$ of the cascade, used e.g. in the DGLAP equations. \\

If the cascade is evolved in the `forward' direction, starting with the cascade initiators and ending with the hard scattering, it is not beforehand possible to decide which partons are spacelike, so this and other complications arise. The problem is solved by adopting the `backward evolution' scheme \cite{ref:backwards-evol}, where we start by choosing the hardest process and work our way back to the shower-initiating partons. The cascade is reconstructed, by making use of the AP splitting kernels  and the $Q$-dependent structure functions (cf. Section~\ref{subsect-ME-PS}). During the evolution, one proceeds to lower and lower values for the virtuality. A minimum limit for the $Q^2$ is set so that the evolution will stop when $Q^2 = Q_{0}^2$, where $Q_{0}^2\sim (1 \mrm{GeV})^2$ \cite{ref:initial-state}.\\

The starting point for the backwards evolution is the DGLAP equation (cf. equation~(\ref{APt}))  
\begin{equation}
\frac{\mrm{d}f_{b}(x,t)}{\mrm{d}t} = \sum_{a} \int_{x}^{1} \frac{\mrm{d}x'}{x'} \, \frac{\as(t)}{2\pi} \,  f_{a}(x',t) \, P_{a\rightarrow bc}(z),
\label{APt-again}
\end{equation}
where, as before, the variable $t=\ln(Q^{2}/\Lambda^{2}_{\mrm{QCD}})$. Now we are working backwards in time, away from the hard scattering. Therefore the DGLAP equation expresses the rate at which partons $b$ of momentum $x=zx'$ are `unresolved' into partons $a$ of momentum $x'$, as we take a step $\mrm{d}t$ backwards. \\
The corresponding relative probability is $\mrm{d}P_{b}/\mrm{d}t = (1/f_{b}) \, (\mrm{d}f_{b}/\mrm{d}t)$. The cumulative effects of many small $\mrm{d}t$ steps are summarized in the so-called Sudakov form factor, which gives the probability that parton $b$ stays at $x$ from $t_{\mrm{max}}$ to $t<t_{\mrm{max}}$:
\begin{eqnarray}
S_{b}(x,t_{\mrm{max}},t) &=& \exp\left( {-\int_{t}^{t_{\mrm{max}}} 
\frac{1}{f_{b}} \, \frac{\mrm{d}f_{b}(x,t')}{\mrm{d}t'} \, \mrm{d}t'} \right)
\nonumber
\\
&=& \exp\left( -\int_{t}^{t_{\mrm{max}}} \mrm{d}t' \sum_{a} \int_{x}^{1}
 \frac{\mrm{d}x'}{x'} \, \frac{\as(t')}{2\pi} \, 
\frac{f_{a}(x',t')}{f_{b}(x,t')} \, P_{a\rightarrow bc}(z) \right).
\end{eqnarray}
The differential probability distribution ${\mrm{d}S_{b}}/{\mrm{d}t}$ that a branching $a\rightarrow bc$ actually takes place between $t$ and $t\!-\!\mrm{d}t$ is
\begin{eqnarray}
\frac{\mrm{d}S_{b}}{\mrm{d}t} 
&=& \sum_{a} \int_{x}^{1}\frac{\mrm{d}x'}{x'} \,  \frac{\as(t)}{2\pi} \,
\frac{f_{a}(x',t)}{f_{b}(x,t)} \, P_{a\rightarrow bc}(z) \times\\
\nonumber
& & \times \exp\left( -\int_{t}^{t_{\mrm{max}}} \mrm{d}t' \sum_{a} \int_{x}^{1}
\frac{\mrm{d}x'}{x'}\, \frac{\as(t')}{2\pi}\, \frac{f_{a}(x',t')}{f_{b}(x,t')} \,
P_{a\rightarrow bc}(z) \right).
\label{dS/dt}
\end{eqnarray}
Once the Sudakov form factor is known, the parton shower can be reconstructed. At each branching, the values for $t$ (also defining the virtuality of parton $b$), $z$ (the splitting variable) and $a$ (the flavour of the mother parton) need to be found. The variable $t$ is chosen according to the differential probability distribution ${\mrm{d}S_{b}}/{\mrm{d}t}$. For a given $t$, $a$ is chosen according to the branching probabilities --- the $x'$ integrals. Finally, with $t$ and $a$ known, the probability distribution for $z$ is given by the mentioned integrand.\\
The process of reconstructing the cascade is iterative; when going from one branching $a\rightarrow bc$ to the next $a'\rightarrow b'c'$ (where $a=b'$), the actual $t$ value of the parton $b$ is taken as the upper limit $t_{\mrm{max}}$ for the parton $b'$. Thus, the virtualities of the partons on the main chain decrease during the evolution. The opposite is true for the momentum fractions $x$. (Remember that we are evolving backwards.)\\

\begin{figure}[htb]
\begin{center}
\begin{picture}(240,100)(0,0)
\LongArrow(10,10)(40,40)
\LongArrow(40,40)(50,75)
\LongArrow(40,40)(80,60)
\Text(60,55)[b]{3}
\LongArrow(80,60)(100,90)
\Text(85,80)[]{4}
\LongArrow(80,60)(120,60)
\Text(100,65)[b]{1}
\LongArrow(230,10)(200,40)
\LongArrow(200,40)(190,75)
\LongArrow(200,40)(160,60)
\Text(180,55)[b]{5}
\LongArrow(160,60)(140,90)
\Text(155,80)[]{6}
\LongArrow(160,60)(120,60)
\Text(140,65)[b]{2}
\end{picture}
\end{center}
\caption{\label{fig:ps-notation} Schematic picture of initial-state parton showers. Partons 1 and 2 take part in the hard scattering.}
\end{figure}
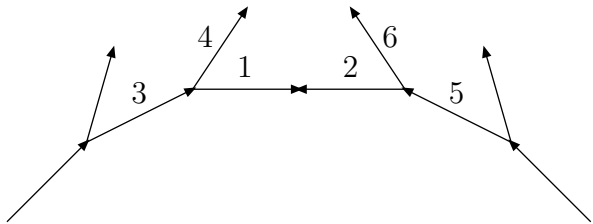

Finally, some words about the interpretation of the variables $z$ and $\hat{s}$. We first consider the case of collinear and massless particles and use the notation of figure~\ref{fig:ps-notation}. 
For the partons 1 and 2 that form the $\mrm{W}$ we have as before
\begin{equation}
(p_1+p_2)^2 = \mtwo{W} (=\hat{s}_{0}).
\label{temp1}
\end{equation}
On the other hand, if parton 1 is unresolved, we have that 
\begin{equation}
(p_3+p_2)^2 =\hat{s}.
\label{temp2}
\end{equation}
The splitting variable $z$ relates the four-momenta of parton 1 and its mother, parton 3, according to
\begin{equation}
p_{1} = z p_{3}.
\end{equation}
By multiplying both sides with $p_{2}$, this can be rewritten as
\begin{equation}
(p_1+p_2)^2 = z \, (p_3+p_2)^2.
\end{equation}
(Here we have used the assumption that the particles are massless.)
From this and equations~(\ref{temp1}) and (\ref{temp2}) we obtain 
\begin{equation}
z = \frac{\mtwo{w}}{\hat{s}}.
\label{z}
\end{equation}
This expression is used iteratively in the `backwards-evolution', thus also for non-collinear emission. We will use this expression as a definition for $z$, when we will make translations between the parton-shower variables $z$ and $Q^2$ and the matrix-elements ones $\hat{s}$, $\hat{t}$ and $\hat{u}$ (cf. Section~\ref{sec-modeling}).



%
\section{Modeling the parton shower according to the matrix element}
\label{sec-modeling}
%
%
%

The parton-shower description has the advantage that one can allow for an arbitrary number of particles. This is in contrast with the matrix-element description, where the calculations become complicated very fast as the number of particles increases. Also, we expect the parton-shower formalism to give a good description of the distribution at low $\pt$ values, and the matrix-element formalism to be useful in the high $\pt$ region.\\  
The main goal of this work is to take the lowest-order matrix element of $\mrm{q}\qbar'\rightarrow \mrm{W}$ with initial-state parton shower, and modify it so that it can be used as an alternative description to the higher-order matrix-element at large $\pt$ values as well.\\

From a QCD point of view, the lowest-order matrix element $\mrm{q}\qbar'\rightarrow \mrm{W}$ is of zeroth order in $\as$. Since we want to describe the $\mrm{W}$ transverse momentum, this `naked' process is uninteresting. Instead we have to allow initial-state radiation. In this Diploma Work, we start by considering the $\mrm{q}\qbar'\rightarrow \mrm{W}$ with initial-state parton shower, and introduce corrections derived from the matrix elements of the processes $\mrm{q}\qbar'\rightarrow \mrm{gW}$ and $\mrm{qg}\rightarrow \mrm{q'W}$. We show that the thus modified parton-shower description can be extended to hold for large $\pt$ values as well. In the rest of this article, the following notation will be used:
\begin{itemize}
  \item PS : $\mrm{q}\qbar'\rightarrow \mrm{W}$ with initial-state parton shower,
  \item ME : $\mrm{q}\qbar'\rightarrow \mrm{gW}$ and $\mrm{qg}\rightarrow \mrm{q'W}$.
\end{itemize}
Experimentally, one might wish to consider the two processes $\mrm{q}\qbar'\rightarrow \mrm{gW}$ and $\mrm{qg}\rightarrow \mrm{q'W}$ together, as they are both of first order in $\as$. We choose to make distinctions between them, and we will see in Section~\ref{sec-results} that this separate treatment is well motivated indeed. Thus, for comparison reasons, we divide the PS into two parts:
\begin{enumerate}
  \item PS similar to $\mrm{q}\qbar'\rightarrow \mrm{gW}$.\\
  The two Feynman diagrams that contribute to this ME are shown in figure \ref{fig:qqbar->gW}. For the PS to be similar to the above ME, the most virtual quark (closest to the hard scattering) must have come from a ${\mrm{q}\rightarrow\mrm{qg}}$ vertex, as is shown in figure~\ref{fig:PS-like-qqbar->gW}.
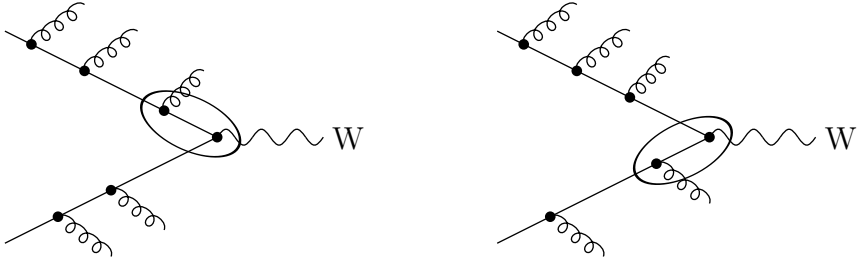
\begin{figure}[tb]
\begin{center}
\begin{picture}(150,120)(0,0)
\Line(10,100)(90,60)
\Gluon(20,95)(40,110){2.5}{3}
\Vertex(20,95){2}
\Gluon(40,85)(60,100){2.5}{3}
\Vertex(40,85){2}
\Gluon(70,70)(85,85){2.5}{3}
\Vertex(70,70){2}
\Line(10,20)(90,60)
\Gluon(30,30)(50,15){2.5}{3}
\Vertex(30,30){2}
\Gluon(50,40)(70,25){2.5}{3}
\Vertex(50,40){2}
\Vertex(90,60){2}
\Photon(90,60)(130,60){3}{3}
\Oval(80,65)(10,20)(155)
\Text(140,60)[]{$\mrm{W}$}
\end{picture}
\hspace{1cm}
\begin{picture}(150,120)(0,0)
\Line(10,100)(90,60)
\Gluon(20,95)(40,110){2.5}{3}
\Vertex(20,95){2}
\Gluon(40,85)(60,100){2.5}{3}
\Vertex(40,85){2}
\Gluon(60,75)(75,90){2.5}{3}
\Vertex(60,75){2}
\Line(10,20)(90,60)
\Gluon(30,30)(50,15){2.5}{3}
\Vertex(30,30){2}
\Gluon(70,50)(90,35){2.5}{3}
\Vertex(70,50){2}
\Vertex(90,60){2}
\Photon(90,60)(130,60){3}{3}
\Oval(80,55)(10,20)(27)
\Text(140,60)[]{$\mrm{W}$}
\end{picture}
\end{center}
\caption{\label{fig:PS-like-qqbar->gW} $\mrm{q}\qbar'\rightarrow \mrm{W}$ with initial-state parton shower. The circled region is similar to the matrix element of $\mrm{q}\qbar'\rightarrow \mrm{gW}$.}
\end{figure}
  \item PS similar to $\mrm{qg}\rightarrow \mrm{q'W}$.\\
  The two Feynman diagrams that contribute to this ME are shown in figure \ref{fig:qg->qW}. For the PS to be similar to the above ME, the most virtual quark (closest to the hard scattering) must have come from a ${\mrm{g}\rightarrow\mrm{q}\qbar}$ vertex, as is shown in figure~\ref{fig:PS-like-qg->qW}.
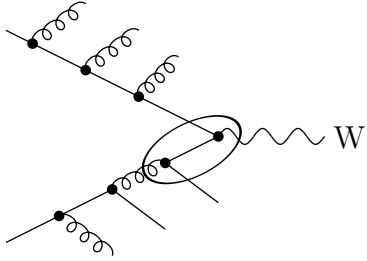
\begin{figure}[tb]
\begin{center}
\begin{picture}(150,120)(0,0)
\Line(10,100)(90,60)
\Gluon(20,95)(40,110){2.5}{3}
\Vertex(20,95){2}
\Gluon(40,85)(60,100){2.5}{3}
\Vertex(40,85){2}
\Gluon(60,75)(75,90){2.5}{3}
\Vertex(60,75){2}
\Line(10,20)(50,40)
\Gluon(30,30)(50,15){2.5}{3}
\Vertex(30,30){2}
\Line(50,40)(70,25)
\Gluon(50,40)(70,50){2.5}{3}
\Vertex(50,40){2}
\Line(70,50)(90,60)
\Line(70,50)(90,35)
\Vertex(70,50){2}
\Vertex(90,60){2}
\Photon(90,60)(130,60){3}{3}
\Text(140,60)[]{$\mrm{W}$}
\Oval(80,55)(10,20)(27)
\end{picture}
\end{center}
\caption{\label{fig:PS-like-qg->qW}  $\mrm{q}\qbar'\rightarrow \mrm{W}$ with initial-state parton shower. The circled region is similar to the matrix element of $\mrm{qg}\rightarrow \mrm{q'W}$.}
\end{figure}
\end{enumerate}
%



%
\subsection{Comparing the ME and PS differential cross sections}
\label{subsec-comparing}
%
%
In this section we compare the differential cross sections for the PS and the ME. An expression for the PS differential cross section is derived from the ME one, as illustrated below.
\begin{enumerate}
  \item $\mrm{q}\qbar'\rightarrow \mrm{gW}$ compared to the corresponding part of the PS.\\
The Feynman diagrams for this ME are shown in figure \ref{fig:qqbar->gW}. The total cross section is given by equation~(\ref{sigma2}) where 
\begin{eqnarray}
\left.\frac{\mrm{d}\hat{\sigma}}{\mrm{d}\hat{t}} \right|_{\mrm{ME}} &=&
\sigma_{0} \, \frac{4}{3} \, \frac{\as}{2\pi} \, \frac{\mtwo{W}}{\hat{s}^2} 
\, \frac{\hat{t}^2+\hat{u}^2+2\mtwo{W}\hat{s}}{\hat{t}\hat{u}}
\nonumber
\\
&=&
\sigma_{0} \, \frac{\as}{2\pi}\,
\frac{\mtwo{W}}{\hat{s}^2} \, \mrm{A}_{\mrm{ME}}.
\label{sigmaME} 
\end{eqnarray}
In order to derive the corresponding differential cross section for the PS, we first rewrite $\mrm{A}_{\mrm{ME}}$, by replacing the matrix-element variables $\hat{s}$, $\hat{t}$ and $\hat{u}$ with the parton-shower variables $z$ and $Q$.\\
The $t$-channel graph in figure \ref{fig:qqbar->gW} is similar to the parton-shower diagram (if only the hardest gluon of the cascade is considered) shown to the left in figure~\ref{fig:PS-like-qqbar->gW}. We can make the following `translations'
\begin{eqnarray}
\hat{s} &=& (p_1+p_2)^2=\frac{\mtwo{W}}{z},
\label{stranslate}
\\
\hat{t} &=& (p_1-p_3)^2=-Q^2,
\label{ttranslate}
\\
\hat{u} &=& \mtwo{W}-\hat{s}-\hat{t}=Q^2-\frac{1-z}{z}\,\mtwo{W}.
\label{utranslate}
\end{eqnarray}       
In the first equation we have used the expression for $z$ of equation~(\ref{z}). In the second one we have equated $\hat{t}$ with the virtuality of the cascade. In the third we have used equation~(\ref{s+t+u}). \\
The expression for $\mrm{A}_{\mrm{ME}}$ can now be rewritten as
\begin{eqnarray}
\mrm{A}_{\mrm{ME}} &=&
\frac{4}{3} \, \frac{\hat{t}^2+\hat{u}^2+2\mtwo{W}\hat{s}}{\hat{t}\hat{u}} 
\nonumber
\\
&=& 
\frac{4}{3} \, \frac{\frac{1+z^2}{z^2} 
\mfour{W}-2\frac{1-z}{z}Q^2\mtwo{W}+2Q^4 } {Q^2(\frac{1-z}{z}\mtwo{W}-Q^2)}
\nonumber
\\
&\approx&
\frac{1}{Q^2} \, \frac{4}{3} \, \frac{1+z^2}{1-z} \, \frac{\mtwo{W}}{z}.
\end{eqnarray}
In the last row we have assumed small virtualities $Q^2\ll \mtwo{W}$, which correspond to the region where PS can be trusted. As expected, we recover the AP splitting kernel $P_{\mrm{q}\rightarrow\mrm{qg}}(z)$ and also the $1/{Q^2}$ factor present in the DGLAP evolution equation (cf. equation~(\ref{APQ})). Thus we see that the parton-shower-formalism expression is indeed recovered at small $\pt$. (The extra factors present in the expression cancel when the kinematics are considered and the integration of equation~(\ref{sigma2}) is performed.)\\
This PS expression for $\mrm{A}$, which we will call $\mrm{A}_{\mrm{PS1}}$, can again be expressed in terms of $\hat{s}$, $\hat{t}$ and $\hat{u}$ by making use of equations (\ref{stranslate}), (\ref{ttranslate}) and (\ref{utranslate}). We thus obtain\\
\begin{equation}
\mrm{A}_{\mrm{PS1}} = \frac{4}{3} \, \frac {\hat{s}^2+\mfour{W}} 
{(\hat{t}+\hat{u}) \hat{t}}.
\\
\end{equation}
The $u$-channel graph in figure \ref{fig:qqbar->gW} is similar to the parton shower diagram (only the hardest gluon of the cascade is considered) shown to the right in figure~\ref{fig:PS-like-qqbar->gW}. The corresponding `translation' equations are obtained as above, simply interchanging $\hat{t}$ and $\hat{u}$. Thus the expression $\mrm{A}_{\mrm{PS2}}$ is  
\begin{equation}
\mrm{A}_{\mrm{PS2}} = \frac{4}{3} \, \frac {\hat{s}^2+\mfour{W}} 
{(\hat{t}+\hat{u}) \hat{u}}.
\nonumber
\\
\end{equation}
Since the parton-shower description is additive (no interference), the total parton-shower answer is simply $\mrm{A}_{\mrm{PS}}=\mrm{A}_{\mrm{PS1}}+\mrm{A}_{\mrm{PS2}}$, i.e.
\begin{eqnarray}
\left.\frac{\mrm{d}\hat{\sigma}}{\mrm{d}\hat{t}} \right|_{\mrm{PS}} 
&=&
\sigma_{0} \, \frac{4}{3}
\frac{\as}{2\pi} \, \frac{\mtwo{W}}{\hat{s}^2} \, \frac{\hat{s}^2+\mfour{W}}
{\hat{t}\hat{u} }
\nonumber
\\
&=&
\sigma_{0} \,
\frac{\as}{2\pi} \, \frac{\mtwo{W}}{\hat{s}^2} \, \mrm{A}_{\mrm{PS}}.
\label{sigmaPS} 
\end{eqnarray}
Finally, we obtain an expression for the ratio $R(\hat{s},\hat{t})$ between the ME and the PS differential cross sections:
\begin{equation}
R(\hat{s},\hat{t})=\frac{ ({\mrm{d}\hat{\sigma}}/{\mrm{d}\hat{t}})_{\mrm{ME}} } { ({\mrm{d}\hat{\sigma}}/{\mrm{d}\hat{t}})_{\mrm{PS}} }= 
\frac{\hat{t}^2+\hat{u}^2+2\mtwo{W}\hat{s}}{\hat{s}^2+\mfour{W}}.
\label{R-qqbar}
\end{equation}
One can show that
\begin{equation}
\frac{1}{2} \leq R(\hat{s},\hat{t}) \leq 1.\\
\label{R-qqbar-up}
\end{equation} 
  \item $\mrm{qg}\rightarrow \mrm{q'W}$ compared to the corresponding part of the PS.\\
The calculations are similar to the previous case. The Feynman diagrams for this ME are shown in figure~\ref{fig:qg->qW}. The total cross section is given by equation~(\ref{sigma2}) where 
\begin{eqnarray}
\left.\frac{\mrm{d}\hat{\sigma}}{\mrm{d}\hat{t}} \right|_{\mrm{ME}} &=&
\sigma_{0} \, \frac{1}{2} \, \frac{\as}{2\pi} \, \frac{\mtwo{W}}{\hat{s}^2} 
\, \frac{\hat{s}^2+\hat{u}^2+2\mtwo{W}\hat{t}}{(-\hat{s})\hat{u}}
\nonumber
\\
&=&
\sigma_{0} \, \frac{\as}{2\pi}\,
\frac{\mtwo{W}}{\hat{s}^2} \, \mrm{A}_{\mrm{ME}}. 
\end{eqnarray}
As before, $\mrm{A}_{\mrm{ME}}$ is to be rewritten by replacing the matrix-element variables $\hat{s}$, $\hat{t}$ and $\hat{u}$ with the parton-shower variables $z$ and $Q$.\\
The $u$-channel graph in figure~\ref{fig:qg->qW} is similar to the parton-shower diagram (if only the hardest gluon of the cascade is considered) shown in figure~\ref{fig:PS-like-qg->qW}. (The $s$-channel graph in figure~\ref{fig:qg->qW} does not have a parton-shower correspondent.) The `translation' equations are similar to equations (\ref{stranslate}), (\ref{ttranslate}) and (\ref{utranslate}), but with the roles of $\hat{t}$ and $\hat{u}$ interchanged. Expressed in the new variables, $\mrm{A}_{\mrm{ME}}$ becomes
\begin{eqnarray}
\mrm{A}_{\mrm{ME}} &=&
\frac{1}{2} \, \frac{\hat{s}^2+\hat{u}^2+2\mtwo{W}\hat{t}}{(-\hat{s})\hat{u}} 
\nonumber
\\
&=& 
\frac{1}{2} \, \frac{  \frac{\mfour{W}}{z^2}+Q^4+2\mtwo{W}Q^2-
2 \left( \frac{1-z}{z}  \right) \mfour{W}  }  {  \frac{\mtwo{W}}{z} Q^2  }
\nonumber
\\
&\approx&
\frac{1}{Q^2} \, \frac{1}{2} \, (z^2+(1-z)^2) \, \frac{\mtwo{W}}{z}.
\end{eqnarray}
In the last row we have again assumed small virtualities $Q^2 \ll \mtwo{W}$, which correspond to the region where PS can be trusted. As expected in this case, we recover the AP splitting kernel $P_{\mrm{g}\rightarrow\mrm{q}\qbar}(z)$ and also the $1/{Q^2}$ factor present in the DGLAP evolution equation (cf.equation~(\ref{APQ})).  
This PS expression for A is rewritten in terms of $\hat{s}$, $\hat{t}$ and $\hat{u}$ 
\begin{equation}
\mrm{A}_{\mrm{PS}} = \frac{1}{2} \, \frac{\hat{s}^2+2\mtwo{W}\hat{t}+ 
2\mtwo{W}\hat{u}} {(-\hat{s})\hat{u}}.
\end{equation}
The parton-shower answer in this case is
\begin{eqnarray}
\left.\frac{\mrm{d}\hat{\sigma}}{\mrm{d}\hat{t}} \right|_{\mrm{PS}} 
&=&
\sigma_{0} \, \frac{1}{2}
\frac{\as}{2\pi} \, \frac{\mtwo{W}}{\hat{s}^2}  \, 
\frac{\hat{s}^2+2\mtwo{W}\hat{t}+ 
2\mtwo{W}\hat{u}} {(-\hat{s})\hat{u}}
\nonumber
\\
&=&
\sigma_{0} \,
\frac{\as}{2\pi} \, \frac{\mtwo{W}}{\hat{s}^2} \, \mrm{A}_{\mrm{PS}}. 
\end{eqnarray}
The expression for the ratio $R(\hat{s},\hat{t})$ between the ME and the PS differential cross sections is thus given by
\begin{equation}
R(\hat{s},\hat{t})=\frac{ ({\mrm{d}\hat{\sigma}}/{\mrm{d}\hat{t}})_{\mrm{ME}} } { ({\mrm{d}\hat{\sigma}}/{\mrm{d}\hat{t}})_{\mrm{PS}} }= 
\frac{\hat{s}^2+\hat{u}^2+2\mtwo{W}\hat{t}}
{\hat{s}^2+2\mtwo{W}\hat{u}+2\mtwo{W}\hat{t}}.
\label{R-qg}
\end{equation}
Again, one can show that
\begin{equation}
1 \leq R(\hat{s},\hat{t}) \leq \frac{\sqrt{5}-1}{2(\sqrt{5}-2)} <  3.\\
\label{R-qg-up}
\end{equation} 
One possible motivation to the fact that $R(\hat{s},\hat{t})$ is bigger in this case is that the $s$-channel graph is absent from the PS description.
\end{enumerate}
%
%


%
\subsection{Description of the modeling}
\label{subsec-description}
%
%

The $\mrm{W}$ transverse momentum distribution is described well by the lowest-order matrix element with parton shower in the region of small transverse momenta, but the description is poor for large ones. This is closely related to the fact that, conventionally, the maximum value for the virtuality associated with the `main chain' in the shower, should be of the order of the hard-scattering scale. In the case where a heavy resonance particle is formed, it is natural to let its mass set the scale for the process. Thus the maximum virtuality is set equal to the square of the mass of the resonance particle, in our case: $Q_{\mrm{max}}^2=\mtwo{W}$. This forces the distribution to fall abruptly for $\pt\sim Q_{\mrm{max}}= \mrm{m_{W}}$. \\

The main goal of this Diploma Work is to modify the PS description of the transverse momentum distribution, so that it can be used as an alternative approach to the higher-order matrix-elements description, also for large $\pt$. The modeling is done by adding corrections derived from the ME.\\
The events are simulated by using the event generator PYTHIA, written in FORTRAN77. In the generator, the initial-state parton shower is implemented as a Monte Carlo simulation. This makes use of the `backwards evolution' scheme (cf. Section~\ref{subsec-back-ev}). The changes that we made have been implemented as a subroutine, which is a modification of the already existing subroutine PYSSPA.\\
The PS is modified in two steps:
\begin{enumerate}
      \item The maximum virtuality is increased.\\
As mentioned above, in the case of a resonance particle being formed, one conventionally sets $Q_{\mrm{max}}^2= \mone{W}^2$. This is also the default value that is already implemented in PYTHIA.\\
On the other hand, when two hadrons are allowed to collide at a c.m. energy of $E_{\mrm{CM}}$, the kinematic limit for the virtuality of the most virtual parton on the `main chain', which actually takes part in the hard scattering, is of the order $E_{\mrm{CM}}^2$. We choose to increase the upper limit for the virtuality, by setting $Q_{\mrm{max}}^2=(E_{\mrm{CM}}/2)^2$.
      \item ME corrections are introduced.\\
The differential probability distribution ${\mrm{d}S_{b}}/{\mrm{d}t}$ of equation (\ref{dS/dt}) is modified, by using the ratio $R(\hat{s},\hat{t})$ between the ME and the PS differential cross sections, derived in Section~\ref{subsec-comparing} (recall that the ratio $R(\hat{s},\hat{t})$ can be rewritten as a function of the parton-shower variables $z$ and $t$). We modify as follows
\begin{eqnarray}
\frac{\mrm{d}\tilde{S}_{b}}{\mrm{d}t} 
&=&  \sum_{a} \int_{x}^{1} \frac{\mrm{d}x'}{x'} \, \frac{\as(t)}{2\pi} \,  
\frac{f_{a}(x',t)}{f_{b}(x,t)} \, P_{a\rightarrow bc}(z)\, R(\hat{s},\hat{t})\, \times
\nonumber
\\
& & \times \, \exp\left( -\int_{t}^{t_{\mrm{max}}} \mrm{d}t' \, \sum_{a} 
\int_{x}^{1} \frac{\mrm{d}x'}{x'} \, \frac{\as(t')}{2\pi} 
\frac{f_{a}(x',t')}{f_{b}(x,t')} \, P_{a\rightarrow bc}(z) \, 
R(\hat{s},\hat{t}) \right).
\label{mod-dS/dt}
\end{eqnarray}
(Remember that the expressions for $R(\hat{s},\hat{t})$ differ for the two different parts of the PS.) This is done only for the branchings closest to the hard scattering, on either side (cf. figures~\ref{fig:PS-like-qqbar->gW} and~\ref{fig:PS-like-qg->qW}). The evolution variable $t$ is chosen according to the corresponding probability distribution $\tilde{S}_{b}(x,t_{\mrm{max}},t)$, by solving the equation
\begin{equation} 
\tilde{S}_{b}(x,t_{\mrm{max}},t)= \int_{t}^{t_{\mrm{max}}}
\frac{\mrm{d}\tilde{S}_{b}}{\mrm{d}t}= R, 
\label{random}
\end{equation}
as usual in a Monte Carlo approach. Here $R$ is a random number between zero and one, chosen from a uniform distribution. To solve equation~(\ref{random}) we make use of the veto algorithm \cite{ref:veto}: the (exact) expression for $R(\hat{s},\hat{t})$ is replaced by $R_{\mrm{max}}(\hat{s},\hat{t})$ (where $R(\hat{s},\hat{t}) \leq R_{\mrm{max}}(\hat{s},\hat{t})$) and $t$ is selected; thereafter the ratio $(R(\hat{s},\hat{t})) / (R_{\mrm{max}}(\hat{s},\hat{t}))$ is used to decide whether to keep this value for $t$, or to evolve further. In the case of the PS part that is similar to $\mrm{q}\qbar'\rightarrow \mrm{gW}$, $R(\hat{s},\hat{t})$ is given by equation~(\ref{R-qqbar}) and $R_{\mrm{max}}(\hat{s},\hat{t})\!=\! 1$ (cf. equation~(\ref{R-qqbar-up})). Analogously, in the case of the PS part that is similar to $\mrm{qg}\rightarrow \mrm{q'W}$, 
$R(\hat{s},\hat{t})$ is given by equation~(\ref{R-qg}) and $R_{\mrm{max}}(\hat{s},\hat{t}) \!= \!3$ (cf. equation~(\ref{R-qg-up})).
\end{enumerate}

In addition to the final runs, intended to show the full complexity of the PS and allow a direct comparison with data, a number of test runs are performed to check the implementation of the ME corrections to the PS.\\ 
In order to check the kinematics of the as-above modified PS, in the framework of the correction factors, we make the PS as similar as possible to the ME. Thus we artificially change the modified PS: we strongly suppress the probabilities of emitting all partons of the shower, except the most virtual (closest to the hard scattering) parton (cf. figure~\ref{fig:PS-like-qqbar->gW} and figure~\ref{fig:PS-like-qg->qW}). (For technical reasons, we can not generate events that are guaranteed to have only one parton in the shower; in our treatment we thus suppress the rate of multiple emissions to very close to zero, and neglect all additional partons.)\\
To isolate the important features, that we want to study and compare for the PS and ME, we choose to generate events in which (1) no associated timelike showers are allowed, i.e. the emitted partons in the initial-state parton shower are put on the mass shell and (2) no final-state radiation is allowed.\\
Finally, we turn to the changes introduced for technical reasons. The main problem, when trying to compare the PS with the ME transverse momentum spectrum for the $\mrm{W}$, is that of too little statistics at large $\pt$. To find a way out of this problem, we first generate events in the usual way (with the above-mentioned changes) to cover low $\pt$ values, and then concentrate on the large $\pt$ region. In the ME case, we generate as above, but impose a minimal $\pt$ value on the generated processes. In the PS case, the situation is somewhat more complicated, and we need to introduce an artificial trick: we generate as above, with the additional requirement that the probability of emitting the most virtual (closest to the hard scattering) parton of the shower is strongly enhanced.  Finally, these spectra are artificially connected at an intermediate $\pt$ value for the PS and the ME, respectively.


%
\section{Results}
\label{sec-results}
%
%

The main topic of this work is the study of the transverse momentum distribution of the $\mrm{W}$ bosons. To do this we generate $\mrm{p}\pbar$ collisions at $\sqrt{s}=1800$ GeV. Experiments involving $\mrm{p}\pbar$ collisions at $\sqrt{s}=1800$ GeV have been conducted at the Fermilab Tevatron Collider.\\ 

\begin{figure}[htb]
\begin{center}
\rotatebox{90}{\mbox{\begin{picture}(260,15)(0,0)
\Text(130,7.5)[]{$\mrm{d}\sigma / \mrm{d}\pt$ (nb/GeV)}\end{picture}}}
{\mbox{\epsfig{file=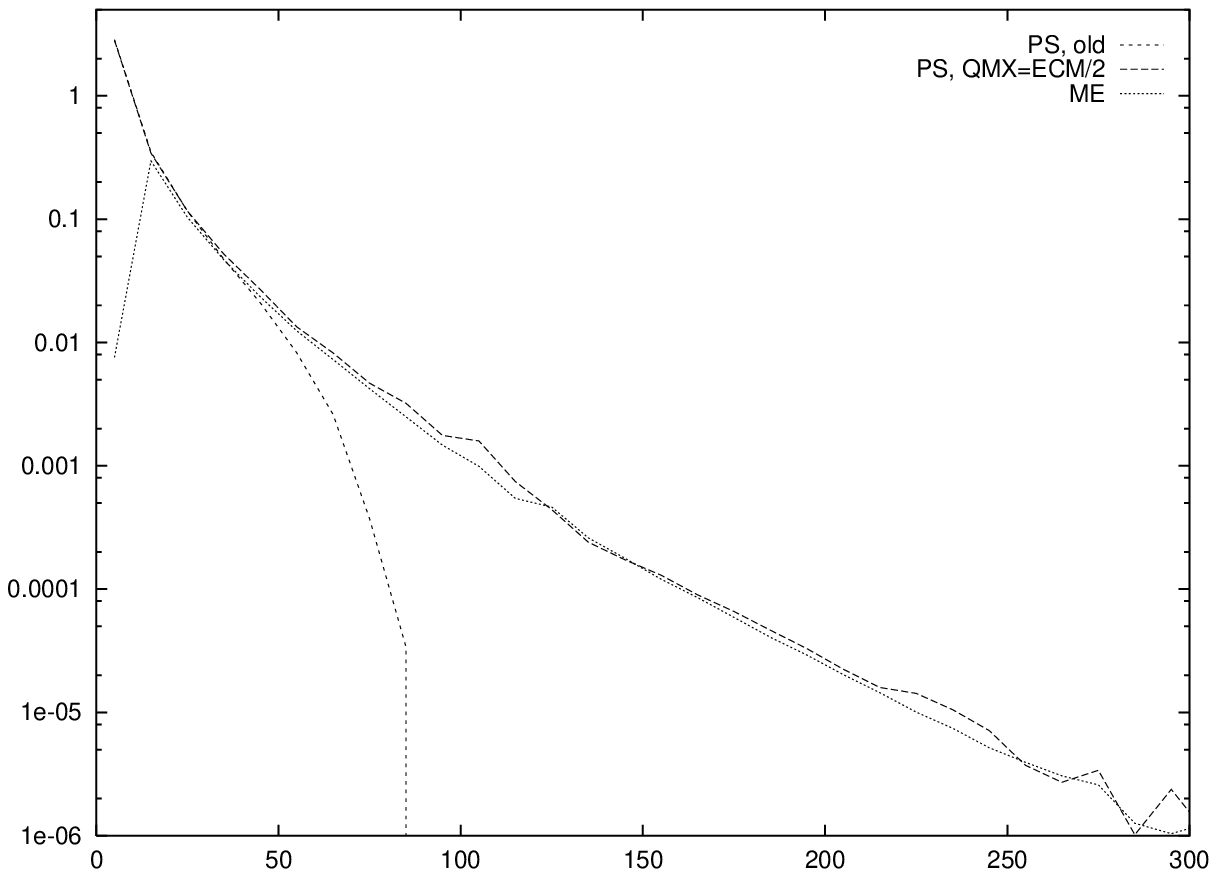}}}
\\
{\mbox{\begin{picture}(400,30)(0,0)
\Text(220,5)[]{transverse momentum for $\mrm{W}$ (GeV)}\end{picture}}}
\end{center}
\captive{\label{fig:1800artif} 
PS (unseparated) compared to the ME. 
}
\end{figure}

\begin{figure}[pt]
\begin{center}
\rotatebox{90}{\mbox{\begin{picture}(260,15)(0,0)
\Text(130,7.5)[]{$\mrm{d}N / \mrm{d}R(\hat{s},\hat{t})$}\end{picture}}}
{\mbox{\epsfig{file=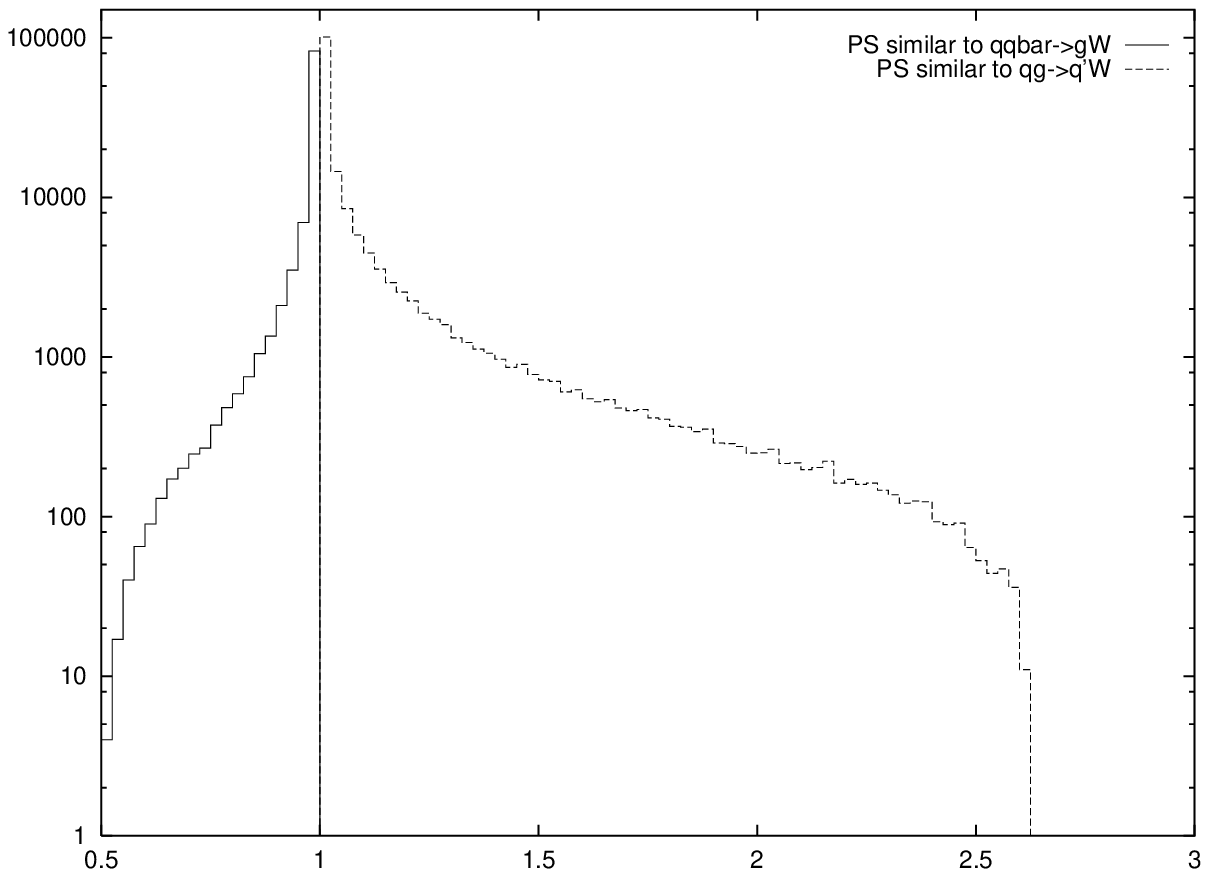}}}
\\
{\mbox{\begin{picture}(400,30)(0,0)
\Text(220,5)[]{$R(\hat{s},\hat{t})$}\end{picture}}}
\end{center}
\captive{\label{fig:R(t)-distr} The $R(\hat{s},\hat{t})$ distributions for the PS similar to  $\mrm{q}\qbar'\rightarrow \mrm{gW}$ and $\mrm{qg}\rightarrow \mrm{q'W}$, respectively.}
\begin{center}
\rotatebox{90}{\mbox{\begin{picture}(260,15)(0,0)
\Text(130,7.5)[]{$\langle R(\hat{s},\hat{t}) \rangle$}\end{picture}}}
{\mbox{\epsfig{file=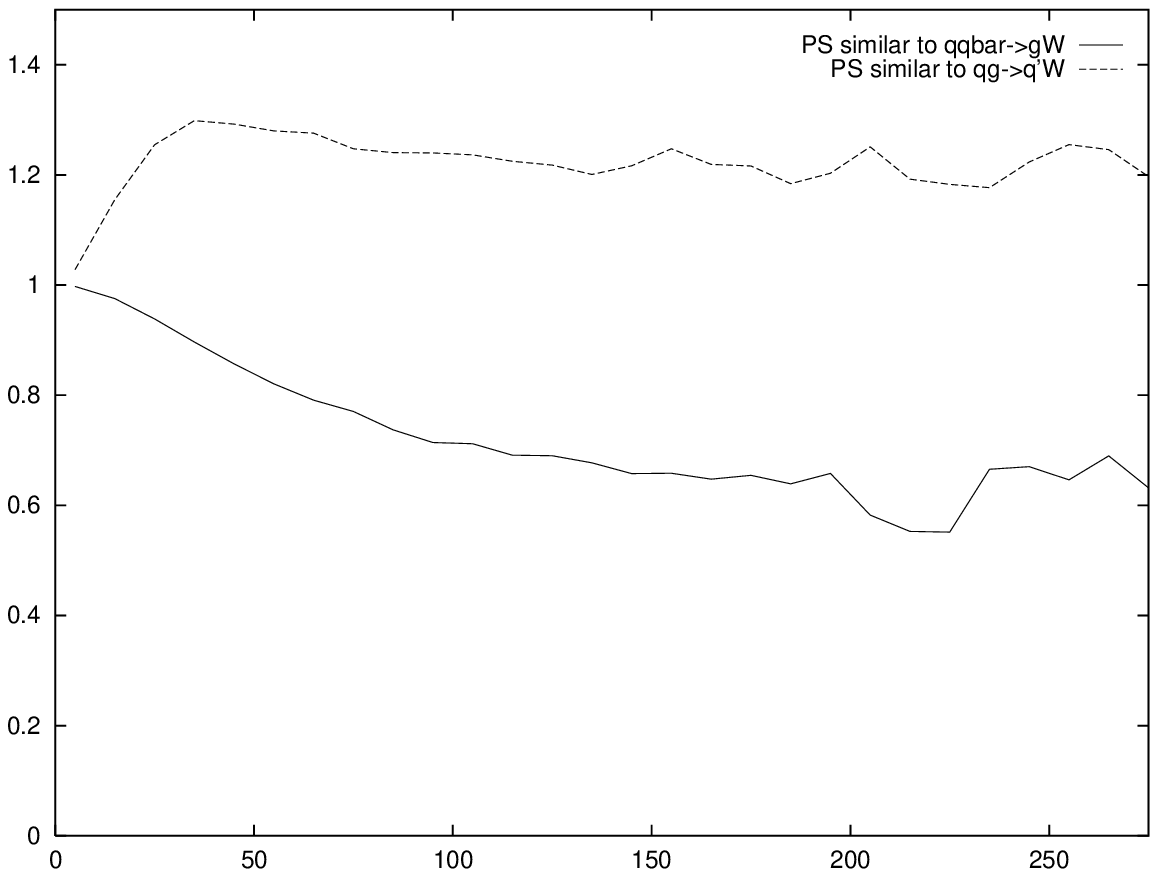}}}
\\
{\mbox{\begin{picture}(400,30)(0,0)
\Text(220,5)[]{transverse momentum for $\mrm{W}$ (GeV)}\end{picture}}}
\end{center}
\captive{\label{fig:<R(t)>} The $\langle R(\hat{s},\hat{t})\rangle $ as a function of the $\mrm{W}$ $\pt$ for the PS similar to  $\mrm{q}\qbar'\rightarrow \mrm{gW}$ and $\mrm{qg}\rightarrow \mrm{q'W}$, respectively.}
\end{figure}

\begin{figure}[p]
\begin{center}
\rotatebox{90}{\mbox{\begin{picture}(260,15)(0,0)
\Text(130,7.5)[]{$\frac{1}{N} \frac{\mrm{d}N}{\mrm{d}\pt}$ (1/GeV)}\end{picture}}}
{\mbox{\epsfig{file=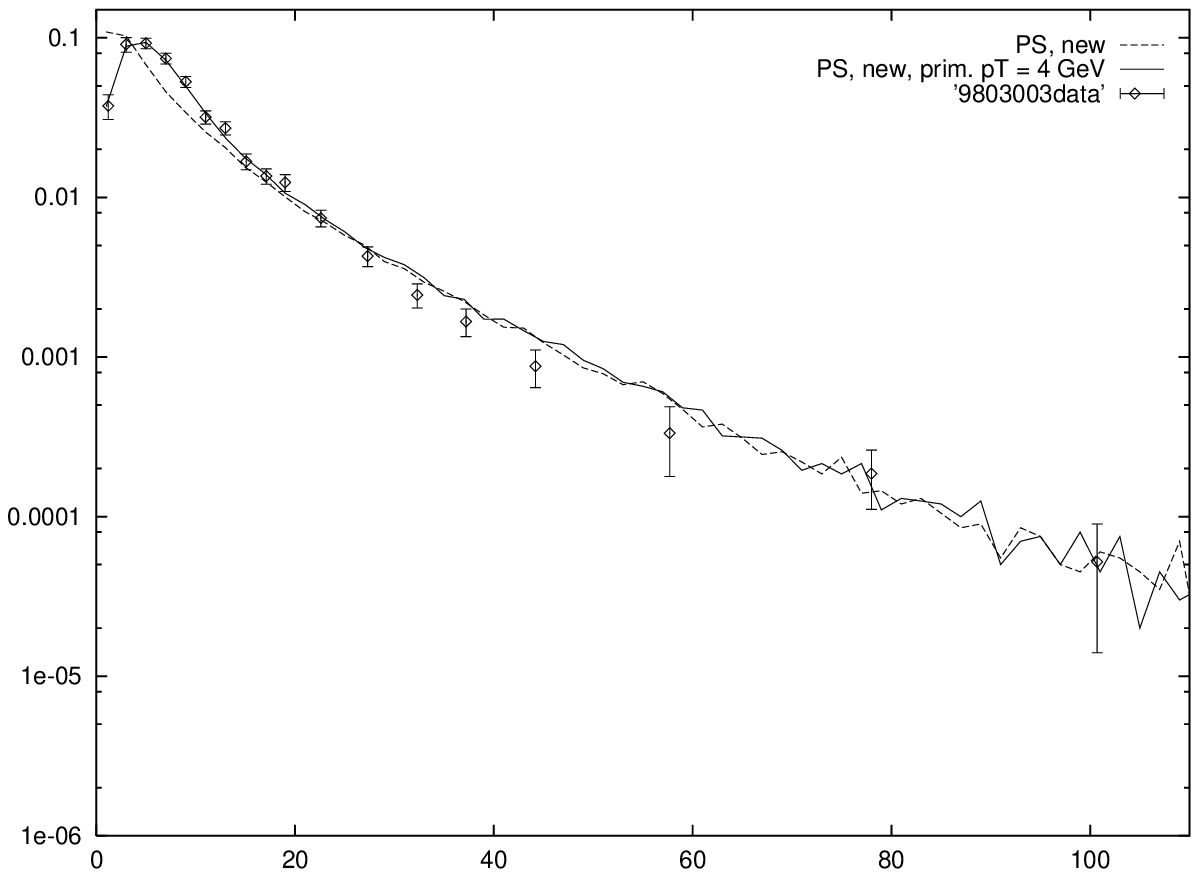}}}
\\
{\mbox{\begin{picture}(400,30)(0,0)
\Text(220,5)[]{transverse momentum for $\mrm{W}$ (GeV)}\end{picture}}}
\end{center}
\captive{\label{fig:1800data} 
The full PS together with data from the D0 collaboration. 
}
\begin{center}
\rotatebox{90}{\mbox{\begin{picture}(260,15)(0,0)
\Text(130,7.5)[]{$\mrm{d}\sigma / \mrm{d}\pt$ (nb/GeV)}\end{picture}}}
{\mbox{\epsfig{file=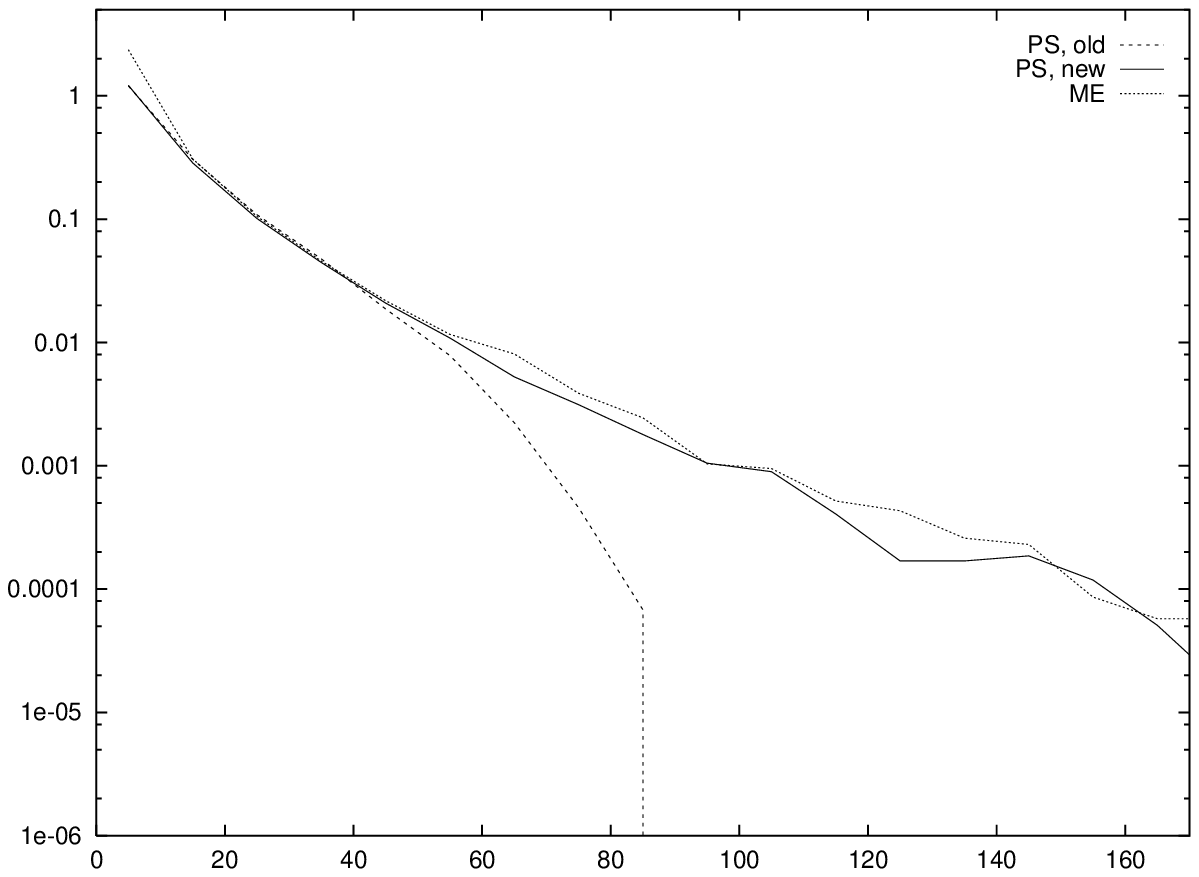}}}
\\
{\mbox{\begin{picture}(400,30)(0,0)
\Text(220,5)[]{transverse momentum for $\mrm{W}$ (GeV)}\end{picture}}}
\end{center}
\captive{\label{fig:1800ME} 
The full PS and the ME. The original PS is also displayed as reference. 
}
\end{figure}

\begin{figure}[htb]{\vspace{-1cm}}
\begin{center}
\rotatebox{90}{\mbox{\begin{picture}(260,15)(0,0)
\Text(130,7.5)[]{(PSnew-PSold)/(PSnew+PSold)}\end{picture}}}
{\mbox{\epsfig{file=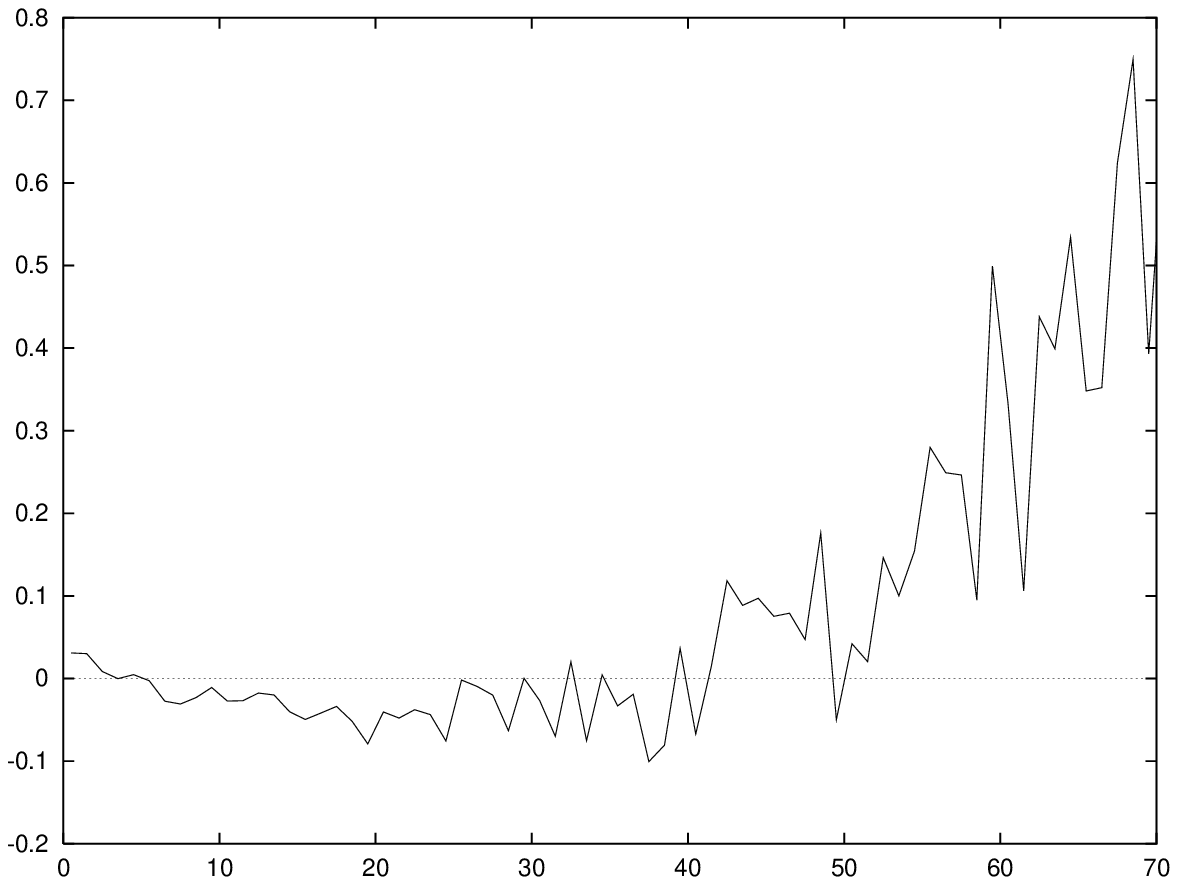}}}
\\
{\mbox{\begin{picture}(400,30)(0,0)
\Text(220,5)[]{transverse momentum for $\mrm{W}$ (GeV)}\end{picture}}}
\end{center}
\captive{\label{fig:1800newold} 
The difference between the full PS and the original PS, normalized to their sum, as a function of the $\mrm{W}$ $\pt$. 
}
\end{figure}

We begin by presenting the test runs, which are performed in order to check the implementation of the ME corrections to the PS. (To do this, the artificial changes mentioned in Section~\ref{subsec-description} are also introduced.)\\
In figure~\ref{fig:1800artif} the (unseparated) PS is compared with the ME. As expected, the distribution of the unchanged PS (short-dashed curve) falls abruptly at $\pt\sim Q_{\mrm{max}}= \mrm{m_{W}}$. Notice that the increase of the maximum virtuality to $Q_{\mrm{max}}^2=(E_{\mrm{CM}}/2)^2$ (long-dashed curve) is enough to prevent this kind of behaviour. Even more, the agreement with the ME (the dotted curve) is indeed very good, already by this simple change. The difference between the ME and the two PS curves, in the low $\pt$ region, is of technical, rather than physical, nature. (The ME has been generated with a $p_{\bot\mrm{min}}=10$ GeV.)\\
\begin{figure}[p]
\begin{center}
\rotatebox{90}{\mbox{\begin{picture}(260,15)(0,0)
\Text(130,7.5)[]{$\mrm{d}\sigma / \mrm{d}\pt$ (nb/GeV)}\end{picture}}}
{\mbox{\epsfig{file=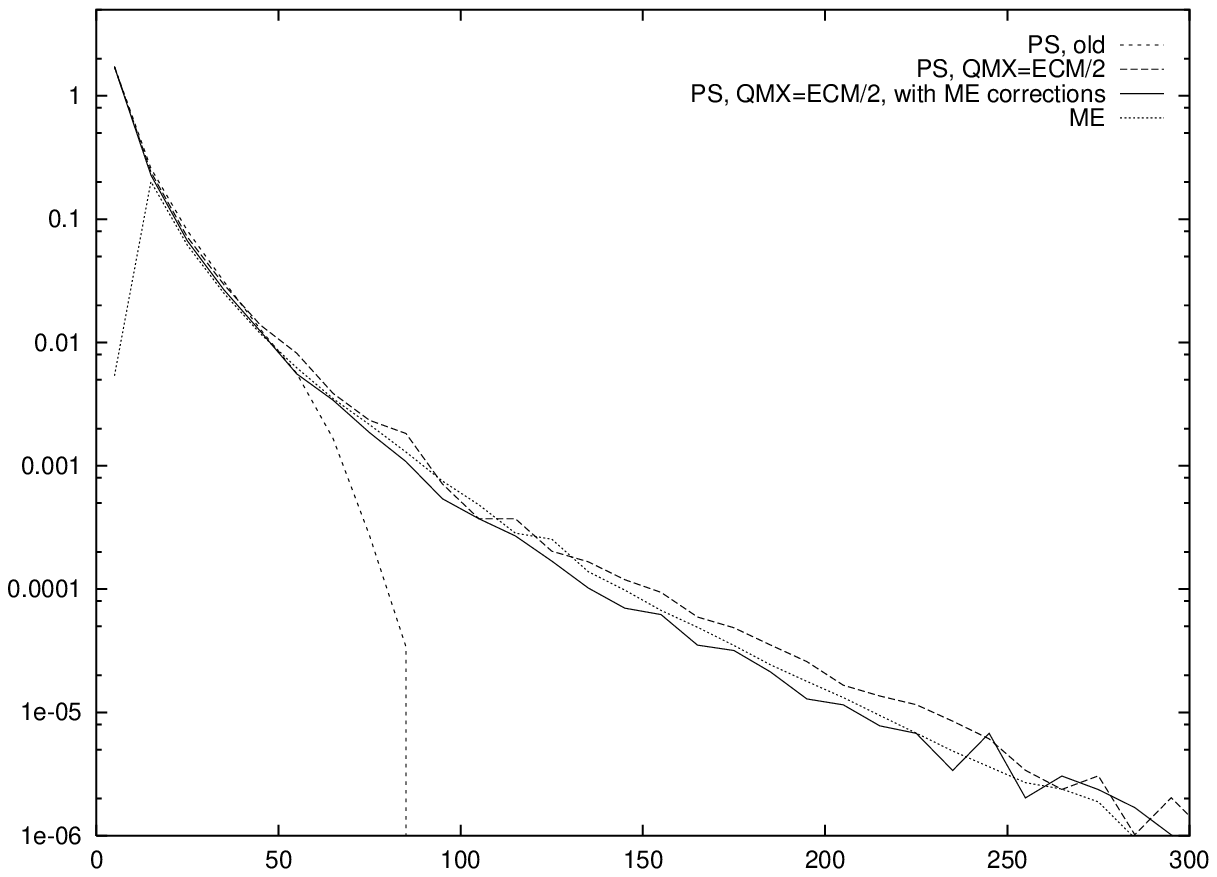}}}
\\
{\mbox{\begin{picture}(400,30)(0,0)
\Text(220,5)[]{transverse momentum for $\mrm{W}$ (GeV)}\end{picture}}}
\end{center}
\captive{\label{fig:qqbar1800artif} 
PS similar to $\mrm{q}\qbar'\rightarrow \mrm{gW}$ compared to the corresponding ME.
}
\begin{center}
\rotatebox{90}{\mbox{\begin{picture}(260,15)(0,0)
\Text(130,7.5)[]{$\mrm{d}\sigma / \mrm{d}\pt$ (nb/GeV)}\end{picture}}}
{\mbox{\epsfig{file=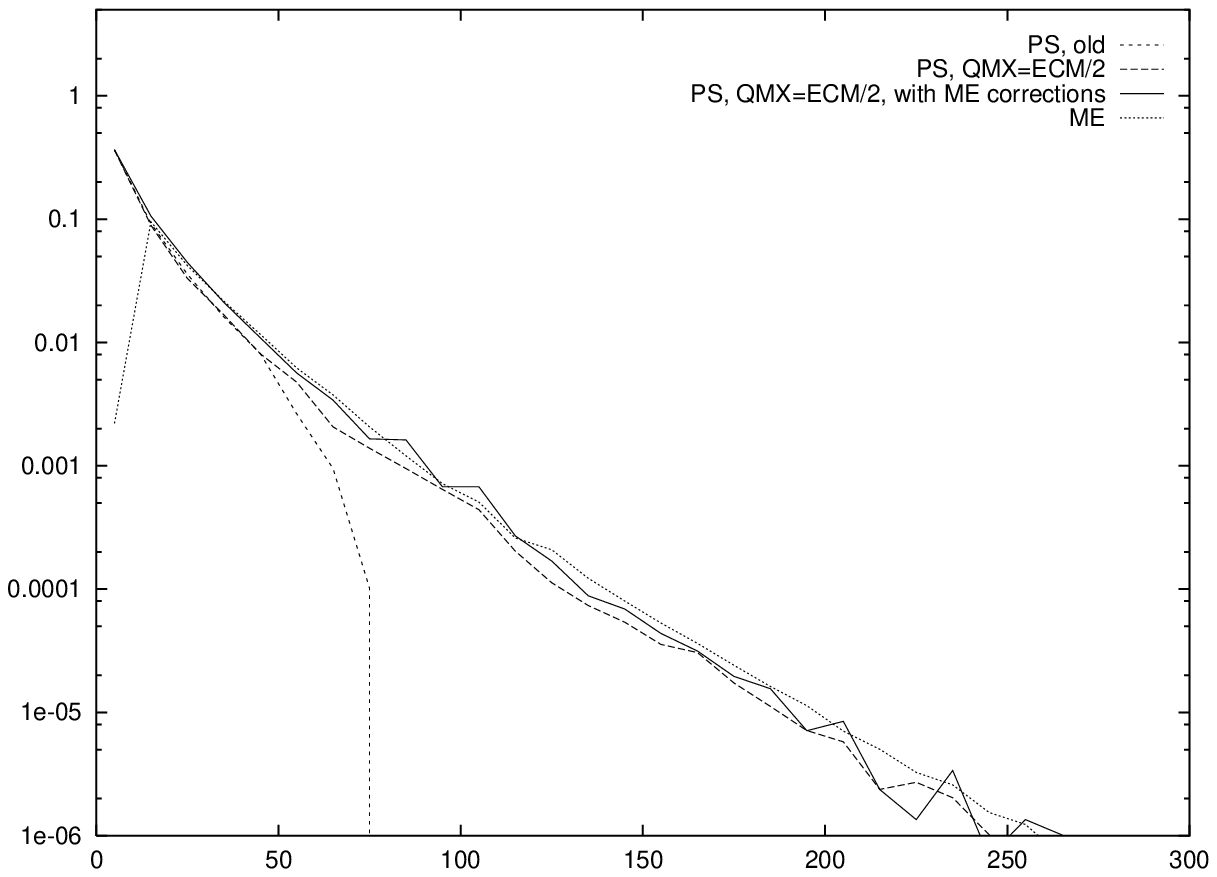}}}
\\
{\mbox{\begin{picture}(400,30)(0,0)
\Text(220,5)[]{transverse momentum for $\mrm{W}$ (GeV)}\end{picture}}}
\end{center}
\captive{\label{fig:qg1800artif} 
PS similar to $\mrm{qg}\rightarrow \mrm{q'W}$ is compared to the corresponding ME. 
}
\end{figure}
Figures~\ref{fig:qqbar1800artif} and \ref{fig:qg1800artif} show the PS similar to $\mrm{q}\qbar'\rightarrow \mrm{gW}$ and to $\mrm{qg}\rightarrow \mrm{q'W}$, respectively, each compared with the corresponding ME (the dotted curves). Just as in the case of the unseparated PS, the distributions of the original PS (short-dashed curves) drop abruptly at $\pt\sim Q_{\mrm{max}}= \mrm{m_{W}}$. After increasing the maximum virtuality (long-dashed curve), the PS of figure~\ref{fig:qqbar1800artif} lies above the ME, while the opposite is true for the PS in figure~\ref{fig:qg1800artif}. (These effects seem to cancel each other out and are therefore not noticeable in figure~\ref{fig:1800artif}.) Clearly, the two parts of the PS need to be treated separately. The corrections derived from the corresponding ME in each case are necessary to give a better matching of the PS to the ME. The full curves include also these corrections. These curves agree quite well with the ME for all $\pt$ values. In the large $\pt$ region, though, they seem to lie systematically slightly below the ME. For reasons of limited time, we have not yet examined this closer.\\
Finally, we consider the  behaviour of the correction factors $R(\hat{s},\hat{t})$. From figure~\ref{fig:R(t)-distr} we conclude that $R(\hat{s},\hat{t})$ is close to unity for most of the events (notice the logarithmic scale), but there are indeed tails of the distributions. Also, the $R(\hat{s},\hat{t})$ values for the two parts of the PS lie in the expected intervals (cf. equations~(\ref{R-qqbar-up}) and~(\ref{R-qg-up})). As can be seen in figure~\ref{fig:<R(t)>}, the $R(\hat{s},\hat{t})\rightarrow 1$ as $\pt \rightarrow 0$, in accordance to what one expects, since it corresponds to the collinear limit. This is also the region with most branchings, which explains the peaked behaviour of the $R(\hat{s},\hat{t})$ distribution. There are variations of the $R(\hat{s},\hat{t})$ for $\pt > 0$, but eventually, for $\pt \gtrsim \mone{W}$, the mean value of the $R(\hat{s},\hat{t})$ becomes fairly constant; this is true for both parts of the PS.\\
\\
We now present the full PS, in all its complexity (the full PS includes the two-steps modification described in Section~\ref{subsec-description}, without any additional changes, unless specified).\\
In figure~\ref{fig:1800data} the full PS (long-dashed curve) is presented together the experimental data \cite{ref:data} from the D0 collaboration (the error bars account for both statistical and systematic uncertainties). The agreement is relatively good, for large $\pt$, but for $\pt$ below $\sim 20$ GeV, the PS is slightly shifted to lower $\pt$ values, compared to the data. To quantify this effect, we also perform runs with a larger primordial $\pt$ than the default value of 0.44 GeV. The distribution corresponding to primordial $\pt=4$ GeV agrees very well with the data. Notice that the increase of the primordial $\pt$ mainly affects the distribution in the low $\pt$ region, leaving it fairly unchanged in the high $\pt$ region.\\
In figure~\ref{fig:1800ME} we present the full PS (short-dashed curve) together with the ME (continuous curve). The distributions agree in the small $\pt$ region; at large $\pt$ we can not really draw any conclusion, since the full PS alternates in being above and below the ME. \\
Finally, in figure~\ref{fig:1800newold}, we compare the full PS with the original one. The plot represents the difference between the full PS and the original PS, normalized to their sum. With the exception of a small region close to zero ($\pt \lessim 5$ GeV), the full PS distribution lies below the original one, up to $\pt \sim 40$ GeV. For larger $\pt$, the full PS lies above the original PS. 
%
%

%
\section{Summary and Outlook}
\label{sec-summary}
%
%
In summary, we conclude that the parton-shower formalism can indeed be extended to hold in the large $\pt$ region as well, by introducing corrections derived from the matrix-element formalism. The parton shower for $\mrm{q}\qbar'\rightarrow \mrm{W}$ has been modified to resemble the matrix elements of $\mrm{q}\qbar\rightarrow \mrm{gW}$ and $\mrm{qg}\rightarrow \mrm{q'W}$ at large $\pt$ values. During the modeling, the separation of the PS into two parts (each corresponding to a matrix element) was necessary. All $\mrm{p}\pbar$ collisions have been performed at $\sqrt{s}=1800$ GeV. When comparing the modified PS with experimental data from the D0 collaboration, we found that the agreement was very good.\\
The parton-shower formalism, modified as described in this article, can be applied in the case of the $\mrm{Z}^{0}$ bosons as well.\\
A possible continuation to this work is to study $\mrm{pp}$ collisions at $\sqrt{s}=14$ TeV. Such collisions will be possible to perform at the LHC at CERN. For the $\mrm{pp}$ runs at $\sqrt{s}=14$ TeV that we have performed (not presented in this article), the discrepancies between the (separated) PS and the corresponding ME are larger than in the case of the $\mrm{p}\pbar$ collisions at $\sqrt{s}=1800$ GeV. We have not yet come so far as to be able to draw any conclusion on whether the differences are due to a kinematic mismatch, or some other physically nontrivial reasons. Once this is overcome, the approach offers hope to model $\pt$ spectra for signal and background processes with good accuracy by rather simple means, so it is of interest to study further. \\

%

%
\subsection*{Acknowledgement}
%
%
%
I would now like to take this opportunity to thank my supervisor, T. Sj\"{o}strand, for his guidance and unlimited patience that he has shown me. I would also like to thank the rest of the staff at the Department of Theoretical Physics, for their help and kind support. Finally, thank you, P\r{a}l, for the encouragement and faith.  

\pagebreak

\end{document}